\documentclass[twocolumn,showpacs,superscriptaddress,aps,prb,amsfonts,amsmath,amssymb,epsf,eqsecnum]{revtex4}
\usepackage{epsfig}
\usepackage[colorlinks]{hyperref}

\newcommand{\ba}{\begin{array}}
\newcommand{\ea}{\end{array}}
\newcommand{\be}{\begin{equation}}
\newcommand{\ee}{\end{equation}}
\newcommand{\nn}{\nonumber}
\newcommand{\bea}{\begin{eqnarray}}
\newcommand{\ena}{\end{eqnarray}}
\newcommand{\beas}{\begin{eqnarray*}}
\newcommand{\enas}{\end{eqnarray*}}
\newcommand{\mb}{\mbox}

\newcommand{\mt}{\mathcal}

\begin{document}

\title{  Characteristics of 2D lattice models from fermionic
realization: Ising  and $XYZ$ models}

\author{Sh. Khachatryan}

\author{A. Sedrakyan}

\affiliation{Yerevan Physics Institute, Br. Alikhanian 2, Yerevan 36, Armenia}

\begin{abstract}
We develop a field theoretical approach to the classical
two-dimensional models, particularly to 2D Ising model (2DIM) and
$XYZ$ model, which is simple to apply for calculation of various
correlation functions. We calculate the partition function of 2DIM
and $XY$ model within the developed framework. Determinant
representation of spin-spin correlation functions is derived using
fermionic realization for the Boltzmann weights. The approach also
allows formulation of the partition function of 2DIM in the
presence of an external magnetic field.
\end{abstract}
 \maketitle
\section{Introduction}

Two-dimensional Ising model (2DIM) \cite{I} is one of the most
attractive models in physics of low dimensions that describe
physical properties of real materials and admit exact solution
\cite{KW,O, K, Y, Kac-Word, KH, MV, Baxter, Chak, PP}. Originally,
2DIM was solved by Onsager \cite{O} in 1944, and, subsequently,
had attracted a steady interest of field theorists and
mathematical physicists. Many effective and interesting approaches
were developed to calculate the free energy, magnetization, and
correlation functions of the model at large distances and all
temperatures. Behavior of the model at the critical point is
governed by the conformal symmetry, and thus, can be well
described by the conformal field theory,, which was developed in
the seminal article by Belavin, Polyakov and Zamolodchikov
\cite{BPZ}. All the critical indices of 2DIM were calculated
within the conformal field theory approach, in full agreement with
the original lattice calculations \cite{O,MV}.

Although various physical characteristics of 2DIM have been
derived using different approaches, still there are open questions
that need to be answered. Some of the most important
characteristics of 2DIM include lattice correlation functions and
form factors \cite{Bug1, Bug2Bug3, Delfino}. These quantities
attract considerable interest in connection with the
condensed-matter problems \cite{Tsvelik-2}, as well as with the
problems in string theory \cite{Polyakov}.
 Importance of form factors becomes especially visible when one
switches on the magnetic field \cite{Delfino, Bha}. Then the
system exhibits the phenomenon, known in particle physics as quark
confinement \cite{Bha}, observed also in spin-1/2 Heisenberg chain
with frustration and dimerization \cite{Tsvelik-2, Aff, Haldane}.

One of the effective approaches to 2DIM is based on its
equivalence to the theory of two-dimensional free fermions (see
Ref. \cite{PP} and references therein) due to the presence of
Kac-Word sign-factor \cite{Kac-Word} in the path integral
representation of the partition function. Though many works have
been dedicated to the investigation of the 2DIM problem by means
of the fermionic (Grassmann) variables, none of them had linked
fermionic representation with vertex $R$ matrix formulation and
possible extensions to other integrable models.

One of the motivations of the present work is to fill this gap and
present a systematically developed field theoretical approach
(action formulation of the partition function) to the 2D Ising and
$XYZ$ models on a square lattice, which is based on the Grassmann
fields. The developed theory utilizes the graded $R$ operator
formalism \cite{S1, S2, KS, AKMS, APSS} and allows the
generalization to other integrable models, which is demonstrated
in this work by operating with rather general $R$ operator.

The paper is organized as follows. In Sec. II first we introduce
the partition function of the 2DIM on the square lattice and
demonstrate that the $R$ matrices, constructed via Boltzmann
weights, satisfy Yang-Baxter equations. Then in Sec. III the
description of fermionic realization for the $R$ matrices is
followed, with particular cases of the eight vertex model, which
is equivalent to the one-dimensional (1D) quantum $XYZ$ model and
2DIM: the case of finite magnetic fields is also considered. In
Sec. IV the partition function is written in the coherent-state
basis in terms of scalar fermions. It is represented as a
continual integral over the fermionic fields with quadratic action
for the 2DIM, when magnetic field vanishes, and for the
free-fermionic limit of the eight-vertex model ( $XY$ model). The
non-local fermionic action is obtained in Sec. IV A for the case
with non-zero magnetic field. Continuum limit of the action is
derived in Sec. IV C. In Sec. IV D the classical results for the
free energy and the thermal capacity are re-obtained within the
developed theory.

In Sec. V we present the technique for fermionic representation of
correlation functions (with details included in the Appendix). In
particular, the two-point correlation functions in 2DIM are
considered on the lattice and their expressions are written in the
Fourier coordinate basis. In the limit of infinite lattice, large
distance spin-spin correlation functions can be presented as a
determinant (Sec. V A), which coincides with the Toeplitz
determinant, studied in Ref. \cite{MV}. Sec. VI is devoted to the
investigation of the spectrum of one-dimensional quantum chain
problem, which is equivalent to the classical 2DIM. The work is
supplemented with an appendix with rather detailed description of
the Jordan-Wigner spin-fermion transformation on 2D lattice, which
we have used in the course of the calculations.

{
\section{Boltzmann weights and Yang-Baxter Equation}}

\paragraph*{\bf{ 1. Boltzmann weights. }}

\begin{figure}[t]
\unitlength=20pt
\leftline{\includegraphics{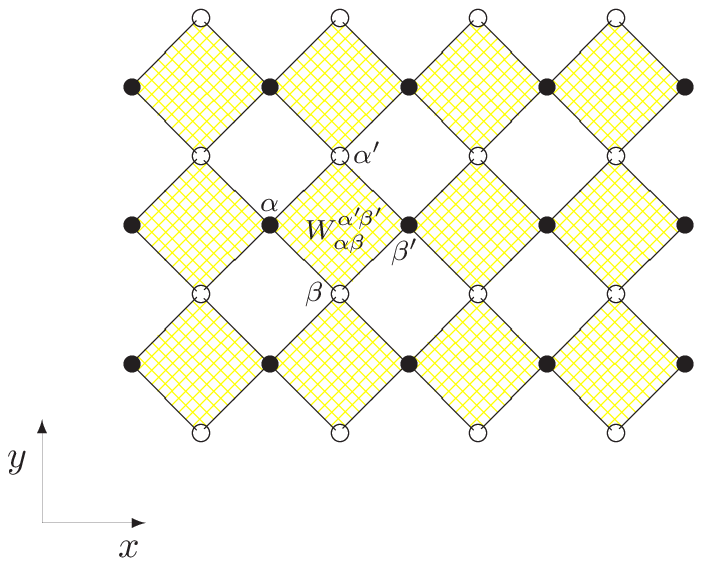}}
%
%
%
%
%
%
%
\caption{A fragment of the lattice of 2DIM: spin variables correspond to
vertices, local Boltzmann weights correspond to
dashed squares. }
\label{fig1}\vspace{0.5cm}
\end{figure}

Classical two-dimensional Ising model on the square lattice can be
defined via its local Boltzmann weights
\bea
 \label{W}
 W_{\alpha\beta}^{\alpha'\beta'}=
e^{J_1(\bar{\sigma}_\alpha\bar{\sigma}_{\alpha'}+\bar{\sigma}_\beta
\bar{\sigma}_{\beta'})+J_2(\bar{\sigma}_\alpha \bar{\sigma}_\beta
+\bar{\sigma}_{\alpha'} \bar{\sigma}_{\beta'})},\\\nn\qquad
\alpha,\;\beta,\;\alpha',\;\beta' =0,1, \ena
where the two state spin variables $\bar{\sigma}_{\alpha}=\{ \pm
1\}$ are assigned to the vertices of the lattice. The partition
function
\be \label{Z} Z(J_1,J_2)=\sum_{\{\alpha,\beta\}}\prod
W_{\alpha\beta}^{\alpha'\beta'} \ee
is a sum over spin configurations of products of the Boltzmann
weights $W_{\alpha\beta}^{\alpha'\beta'}$,  each associated with
the elementary square plaquette with vertices
$\alpha,\;\beta,\;\alpha',\;\beta'$ and arranged in a checkerboard
pattern (dashed squares in Fig.\ref{fig1}). There are imposed
periodic boundary conditions on the spin variables.

Boltzmann weight $W_{\alpha\beta}^{\alpha'\beta'}$ in Eq.
(\ref{W}) can be regarded as a matrix,
\be \label{W2}
W=\left( \ba{cccc} e^{2(J_1+J_2)}&1&1&e^{2(-J_1+J_2)}\\
1&e^{2(J_1-J_2)}&e^{-2(J_1+J_2)}&1\\
1&e^{-2(J_1+J_2)}&e^{2(J_1-J_2)}&1\\
e^{2(-J_1+J_2)}&1&1&e^{2(J_1+J_2)}\\
\ea \right),
\ee
acting as a linear operator on the direct product of two
two-dimensional linear vector spaces,
\be|\alpha\rangle \;|\beta\rangle,\quad \alpha,\; \beta=0,1;\qquad
\langle\beta'|\;\langle\alpha'|\;W\;|\alpha\rangle
\;|\beta\rangle=W_{\alpha\beta}^{\alpha'\beta'},\ee
where $ |0\rangle,\;|1\rangle$ are orthonormalized vectors
$\left(\; _1^0\;\right),\;\left(\; _0^1\;\right)$. The matrix
(\ref{W2}) can be represented as a tensor product of spin
operators,
\bea \label{W1} &\textbf{W}=\frac{e^{2(J_1+J_2)}}{2}({\hat
1}\otimes{\hat 1}+\sigma_z\otimes\sigma_z)&\\\nn&+
\frac{e^{2(J_1-J_2)}}{2}({\hat 1}\otimes{\hat
1}-\sigma_z\otimes\sigma_z)
+\frac{e^{2(-J_1+J_2)}}{2}(\sigma_1\otimes\sigma_1-\sigma_2\otimes\sigma_2)&\\\nn&+
\frac{e^{-2(J_1+J_2)}}{2}(\sigma_1\otimes\sigma_1+\sigma_2\otimes\sigma_2)
+({\hat 1}\otimes\sigma_1+\sigma_1\otimes{\hat 1}),&\ena
%
where $\sigma_\alpha ( \alpha=1,2,z) $ are Pauli matrices and
${\hat 1}$ is the two-dimensional identity operator.
By use of a unitary transformation, one can represent the matrix
$W$ in the form of the $R$ matrix, corresponding to the eight
vertex model. Let us define the unitary matrix
$U=\frac{1}{\sqrt{2}}\left({ }^{\;1}_{\;1}\;
{}^{-1}_{\;\;\;1}\right)$, such that
\newline
%
 $U\sigma_z U^{-1}=\sigma_1,\qquad U\sigma_1
U^{-1}=\sigma_z,\qquad U\sigma_2 U^{-1}=-\sigma_2.$
%
Then the action of these unitary transformations on the linear
spaces, which are assigned to every site of the two-dimensional
square lattice (Fig.\ref{fig1}), yields
%
\bea\nn&  \textbf{R}=\left(U^{-1}\otimes
U^{-1}\right)\textbf{W}\left(U\otimes
U\right)=\frac{e^{2(J_1+J_2)}}{2}({\hat 1}\otimes{\hat
1}+\sigma_1\otimes\sigma_1)&\\\nn&+ \frac{e^{2(J_1-J_2)}}{2}({\hat
1}\otimes{\hat 1}-\sigma_1\otimes\sigma_1)
+\frac{e^{2(J_2-J_1)}}{2}(\sigma_z\otimes\sigma_z-\sigma_2\otimes\sigma_2)&\\\label{R1}&+
\frac{e^{-2(J_1+J_2)}}{2}(\sigma_z\otimes\sigma_z+\sigma_2\otimes\sigma_2)
+({\hat 1}\otimes\sigma_z+\sigma_z\otimes{\hat 1}).&\ena
%
Partition function (\ref{Z}) of the model can be expressed via new
weights (\ref{R1}), as
\bea \label{z} Z=\sum_{\{\alpha,\beta\}}\prod
R_{\alpha\beta}^{\alpha'\beta'}, \ena
where $R$ operator has the following matrix form
\begin{widetext}
{
 \bea \label{rr}
R=2\left(\!\!\! \ba{cccc} \!\cosh{[2J_1]}\cosh{[2J_2]}+1&0&0&\cosh{[2J_1]}\sinh{[2J_2]}\!\\
0&\!\sinh{[2J_1]}\cosh{[2J_2]}&\sinh{[2J_1]}\sinh{[2J_2]}\!&0\\
0&\!\sinh{[2J_1]}\sinh{[2J_2]}\!&\!\sinh{[2J_1]}\cosh{[2J_2]}\!&0\\
\!\cosh{[2J_1]}\sinh{[2J_2]}\!&0&0&\!\cosh{[2J_1]}\cosh{[2J_2]}-1\!\\&
 \ea\!\!\! \right).\ena}
\end{widetext}
>From (\ref{rr}) it is apparent that
$R_{\alpha\beta}^{\alpha'\beta'}$ has the form of $R$ matrix
corresponding to the XY model. It fulfills the "free-fermionic"
condition of the XY model:
\be
R_{00}^{00}R_{11}^{11}-R_{00}^{11}R_{11}^{00}=R_{01}^{01}R_{10}^{10}-R_{01}^{10}R_{10}^{01}.
\label{free} \ee

\paragraph*{\bf{2. Yang-Baxter equations. }}

In this section we examine whether matrix (\ref{rr}) is a solution
of Yang-Baxter equations. We shall verify this by using Baxter's
transformation \cite{Baxter},
\bea &e^{\pm 2J_2}=\mathrm{cn}[i \;u,k]\mp i \;\mathrm{sn}[i\;u,k],&\nn \\
&e^{\pm 2J_1}=i (\mathrm{dn}[i \;u,k]\pm 1)/(k
\;\mathrm{sn}[i\;u,k]).&\label{tranB} \ena
It has been proven in Ref. \cite{Baxter}, that for fixed $k$
parameter, two transfer matrices with different parameters $u$
commute. The case of $k=1$ corresponds to the point of phase
transition.

Now we rewrite the $R$ matrix (\ref{rr}) in terms of functions
 (\ref{tranB}),
%
\begin{widetext}
\bea\label{r}
 R(u,k)=\left(\ba{cccc}
 1+i \frac{\mathrm{cn}[i\; u,\;k]\;\mathrm{dn}[i\;u,\;k]}
 {k\;\mathrm{sn}[i\;u,\;k]}&0&0&\frac{\;\mathrm{dn}[i\;u,\;k]}{k}\\
0&i\frac{\mathrm{cn}[i\;u,\;k]}{k\;\mathrm{sn}[i\;u,\;k]}&\frac{1}{k}&0\\
0&\frac{1}{k}&i\frac{\mathrm{cn}[i\;u,\;k]}{k\;\mathrm{sn}[i\;u,\;k]}&0\\
\frac{\;\mathrm{dn}[i\;u,\;k]}{k}&0&0&-1+i \frac{\mathrm{cn}[i\;
u,\;k]\;\mathrm{dn}[i\;u,\;k]}{k\;\mathrm{sn}[i\;u,\;k]}
\ea\right).
 \ena
 \end{widetext}
Let us multiply matrix (\ref{r}) by $-i\;k\;\mathrm{sn}(i\;u,k)$
and define the matrix ${\bf{r}}(u,k)$ as
\be {\bf{r}}(u,k)=-i\;k\;\mathrm{sn}[i\;u,k] R(u,k). \ee
It is straightforward to verify that ${\bf{r}}(0,k)=I$, where
$I={\hat 1} \otimes{\hat 1}$  is the identity matrix. Importantly,
it takes place the relation
\be{\bf{r}}(-u,k)={\bf{r}}^{-1}(u,k). \ee

Using the properties of the Jacobi elliptic functions, one can
verify that ${\bf{r}}(u,k)$ satisfies the Yang-Baxter equation
\bea\nn\sum_{\beta_1,\;\beta_2,\;\beta'_2}
{\bf{r}}_{\alpha_1\alpha_2}^{\beta_1\beta_2}(u-v,k)
{\bf{r}}_{\beta_2\alpha_3}^{\beta'_2\gamma_3}(u,k)
{\bf{r}}_{\beta_1\beta'_2}^{\gamma_1\gamma_2}(v,k)\quad\;\;\;\;\;=\\
\sum_{\beta_2,\;\beta'_2,\;\beta_3}
{\bf{r}}_{\alpha_2\alpha_3}^{\beta_2\beta_3}(v,k)
{\bf{r}}_{\alpha_1\beta_2}^{\gamma_1\beta'_2}(u,k)
{\bf{r}}_{\beta'_2\beta_3}^{\gamma_2\gamma_3}(u-v,k).\;\;\;\;\;\ena

Note that there is also another $R$ matrix corresponding to 2DIM.
It is known, that the classical 2DIM is a special case of the
eight-vertex model \cite{Baxter}. In general, the $R$ matrix of
the eight-vertex  (or $XYZ$) model can be parameterized by two
model parameters, $k$ and $ \lambda$, as
\begin{widetext}
\bea\label{rxyz}
 r_{xyz}\!=\!\!\left(\!\!\ba{cccc}
  \frac{\mbox{sn}[i\frac{\lambda-u}{2},\;k]}{\mbox{sn}[i\lambda,\;k]}&0&0&-{
  k\;\mbox{sn}[i\frac{\lambda+u}{2},k]\mbox{sn}[i\frac{\lambda-u}{2},k]}\\
0&1&\frac{\mbox{sn}[i\frac{\lambda+u}{2},\;k]}{\mbox{sn}[i\;\lambda,\;k]}&0\\
0&\frac{\mbox{sn}[i\frac{\lambda+u}{2},\;k]}{\mbox{sn}[i\;\lambda,\;k]}&1&0\\
-{k\;\mbox{sn}[i\frac{\lambda+u}{2},k]\mbox{sn}[i\frac{\lambda-u}{2},k]}
&0&0&\frac{\mbox{sn}[i\frac{\lambda-u}{2},\;k]}{\mbox{sn}[i\;\lambda,\;k]}
\ea\!\!\right).
 \ena
\end{widetext}
  The "Ising" limit
corresponds to the choice of $\lambda=\frac{1}{2}I'$, where  $I'$
is one of two half-periods of the elliptic functions
\cite{Baxter}. In this case the eight-vertex model, defined on a
rectangular lattice, splits into two independent Ising models
defined on the two sublattices.

\paragraph*{\bf 3. The transfer matrix and the Hamiltonian. }

\begin{figure}[ht]
\unitlength=20pt
\leftline{\includegraphics{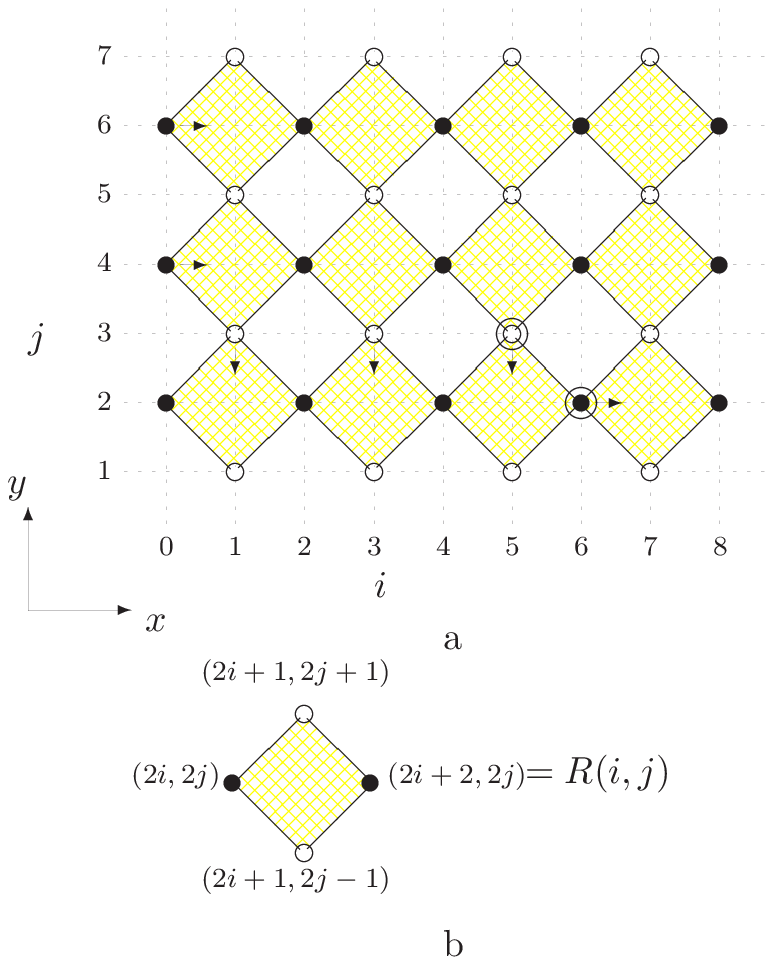}}
\caption{(a) Bold lines and circled vertices represent a fragment
of the lattice of the model in coordinate plane (dotted lines).
(b) Local $R$ operator. }\label{fig2}
\end{figure}

For convenience we denote the coordinates of the lattice sites by
even-even $(2i,2j)$ (black circles on Figs. \ref{fig1} and
\ref{fig2}), and odd-odd  $(2i+1,2j+1)$ (white circles on Figs.
\ref{fig1} and \ref{fig2}) numbers, and assign two-dimensional
linear spaces of quantum states of spins $|\alpha_{2i,2j}\rangle$
and $|\alpha_{2i+1,2j+1}\rangle$ to each of these spaces. Periodic
boundary conditions imply
\be|\alpha_{0,2j}\rangle=|\alpha_{2N,2j}\rangle\quad \mb{and}\quad
|\alpha_{2i+1,1}\rangle=|\alpha_{2i+1,2N+1}\rangle.\label{perbc}\ee

Local $R(i,j)$ operators (\ref{rr}) are acting linearly on the
product of spaces $|\alpha_{2i,2j}\rangle
|\alpha_{2i+1,2j-1}\rangle$ at the sites $(2i, 2j)$ and $(2i+1,
2j-1)$, moving them onto the sites $(2i+1, 2j+1)$ and $(2i+2,
2j)$,
\be R(i,j):|\alpha_{2i,2j}\rangle \; |\alpha_{2i+1,2j-1}\rangle
\Rightarrow|\alpha_{2i+1,2j+1}\rangle \;
|\alpha_{2i+2,2j}\rangle.\ee

%
%
The product of $R$ matrices on each chain along $x$ direction on
the lattice (see Fig.\ref{fig2}) in the formulas of partition
function (\ref{z}), after summation over the boundary states,
constitutes a transfer matrix, $\tau$. Taking into account
conditions (\ref{perbc}), we obtain the following representation
for the transfer matrices
\be \label{tr1} \tau_{j}=\mathrm{tr}_1 \prod_{i=N-1}^{\;0}
R(i,j)\equiv\sum_{\alpha_{0,2j}}\langle\alpha_{0,2j}|\prod_{i=N-1}^{\;0}
R(i,j)\;|\alpha_{0,2j}\rangle. \ee
They act on the states of spins at sites $\{ 2i+1,2j-1
\}_{i=0,N-1}$,
$$|\Sigma_j\rangle=|\alpha_{1,2j-1}\rangle|\alpha_{3,2j-1}\rangle\cdots
|\alpha_{2N-1,2j-1}\rangle,$$ and map them onto the states at
sites $\{ 2i+1,2j+1 \}_{i=0,N-1}$,
$$|\Sigma_{j+1}\rangle=|\alpha_{1,2j+1}\rangle|\alpha_{3,2j+1}\rangle\cdots
|\alpha_{2N-1,2j+1}\rangle.$$

One can interpret the $y$ direction, marked by integers $j$, as a
time direction, while the transfer matrix $\tau_{j}$ will be the
evolution operator for discrete lattice time. In terms of the
transfer matrices,  partition function (\ref{z}) acquires the
following form:
\be \label{tr2}
 Z=\mathrm{tr}_2 \prod_{j=N}^{1} \tau_{j}\equiv
 \sum_{\{\alpha_{2i+1,1}\}_{i=0,N-1}}\langle\Sigma_{1}|\;\prod_{j=N}^{1}
 \tau_{j}\;\;|\Sigma_1\rangle.
 \ee
In Eqs. (\ref{tr1} and \ref{tr2})  $\mathrm{tr}_1$ and
$\mathrm{tr}_2$ represent sums of the states at the boundaries
with (even, even) and (odd, odd) coordinates (dark and light
circles in the figures), respectively. Borrowing  the usual
terminology of the transfer-matrix theory, we can refer the states
marked by white circles on the lattice as "quantum" states, while
the states marked with black circles as "auxilary" states.
\paragraph*{\bf 4. 2D classical model as a $(1+1)$D quantum theory.}

Following to Ref. \cite{C}, one can introduce the limit
\be J_1\sim J \Delta t,\qquad e^{-J_2}\sim h \Delta t, \qquad
\Delta t \ll 1 \label{trot}
\ee
for the continuous time,  in order to establish connection between
two-dimensional classical Ising model and a {\em quantum}
one-dimensional model. In this limit $R$ matrix (\ref{rr}) becomes
\be
 R=\frac{I}{h\Delta t} +\left( \ba{cccc} 1&0&0&J/h\\
0&0&J/h&0\\
0&J/h&0&0\\
J/h&0&0&-1\\\ea \right)+\frac{h \Delta t}{2}\left( \ba{cccc} 1&0&0&0\\
0&-1&0&0\\
0&0&-1&0\\
0&0&0&1\\\ea \right).
\ee
It acquires the following operator form:
\bea \textbf{R}=\!\frac{1}{h\Delta
t}\left\{\hat{1}\otimes\hat{1}\!+\!2\Delta t\left(J
\sigma_1\otimes\sigma_1 \!+\! h (1\otimes\sigma_z\! +\!
\sigma_z\otimes 1)\right)\!\right\}\!\!. \;\;\;\;\;\label{rt}\ena
The coefficient of $\Delta t$ in $\log{\tau}$, constructed with
$R$-matrices (\ref{rt}), defines 1D quantum Hamiltonian for the
Ising model on a chain in a {\em transverse} magnetic field $h$
\cite{Chak},
\be {\cal H}=\sum_i\left(J \sigma_1(i)\sigma_1(i+1) +h
\sigma_z(i)\right).\ee
\paragraph*{\bf 5. Finite magnetic field.}
The Boltzmann weights of the classical 2DIM in a uniformly applied
magnetic field $B$ have the following matrix representation:
 $
 {W_B\:}_{\alpha\;\beta}^{\alpha'\beta'}=
e^{J_1(\bar{\sigma}_\alpha\bar{\sigma}_{\alpha'}+\bar{\sigma}_\beta
\bar{\sigma}_{\beta'})+J_2(\bar{\sigma}_\alpha \bar{\sigma}_\beta
+\bar{\sigma}_{\alpha'} \bar{\sigma}_{\beta'})+\frac{B}{2}
(\bar{\sigma}_\alpha +\bar{\sigma}_\beta +\bar{\sigma}_{\alpha'}
+\bar{\sigma}_{\beta'})}. $
Operator representation of  $W_B$ 
 after unitary
transformation gets 
\bea \label{Rb}&\textbf{R}_B=U^{-1}\otimes U^{-1}\textbf{W}_B U
\otimes U&\\\nn &=\frac{e^{2(J_1+J_2)}}{2}\cosh{[2B]}({\hat
1}\otimes{\hat 1}&\\\nn
&+\sigma_1\otimes\sigma_1)+\frac{e^{2(J_1-J_2)}}{2}({\hat
1}\otimes{\hat 1}-\sigma_1\otimes\sigma_1) &\\\nn
&+\frac{e^{2(J_2-J_1)}}{2}(\sigma_z\otimes\sigma_z-\sigma_2\otimes\sigma_2)&\\\nn
&+
\frac{e^{-2(J_1+J_2)}}{2}(\sigma_z\otimes\sigma_z+\sigma_2\otimes\sigma_2)&\\\nn
&+\frac{e^{2(J_1+J_2)}}{2}\sinh{[2B]}({\hat
1}\otimes\sigma_1+\sigma_1\otimes{\hat 1})&\\\nn &
+\sinh{[B]}(\sigma_1\otimes\sigma_z+\sigma_z\otimes\sigma_1)
+\cosh{[B]}({\hat 1}\otimes\sigma_z+\sigma_z\otimes{\hat 1}).&
\ena
%
In the limit
\be J_1\sim J \Delta t,\qquad e^{-J_2}\sim h \Delta t,\qquad B\sim
\mt{B}\Delta t, \qquad \Delta t \ll 1, \ee
operator $R_B$ obtains the form
\bea\nn \textbf{R}_B&=&\frac{1}{h\Delta t}\left\{{\hat
1}\otimes{\hat 1}+2\Delta t\left(J \sigma_1\otimes\sigma_1 +h
({\hat
1}\otimes\sigma_z+\sigma_z\otimes {\hat 1})\right.\right.\\
&+&\left.\left.\frac{\mt{B}}{2}({\hat
1}\otimes\sigma_1+\sigma_1\otimes {\hat 1})\right)\right\}, \ena
which establishes an equivalence with the quantum 1DIM in the
 magnetic field $\mt{B}$. In this case
 the operator $\frac{1}{\Delta t}\log{\tau}$, when $\Delta t\to
 0$, defines the following structure
 of Hamiltonian operator
$$ {\cal H_B}=\sum_i\left(J
\sigma_1(i)\sigma_1(i+1) + h
\sigma_z(i)+\frac{\mt{B}}{2}\sigma_{1}(i)\right).$$
%
{\centering \section{Fermionic realization of Boltzmann weights}}

Besides the matrix formulation for Boltzmann weights (\ref{W1})
and (\ref{R1}), one can think of alternative representations.
Examples include representation of the $R$ matrix in the Fock
space of the scalar fermions and in a space with the basis of
fermionic coherent-states, developed in Refs. \cite{S1, S2, APSS,
KS}. These reformulations are fully equivalent, and they allow
developing a field theory corresponding to the model. The latter
simplifies calculations of physical quantities (particularly, the
free energy and the magnetization) of the model, and can be
extended for the computation of form factors too, which are
problematic in the standard scheme.

Let us now consider the graded Fock space of scalar fermions
$c^+({i,j})$ and $c({i,j})$, on the lattice,
$([c^+({i,j}),c({i,j})]_{+}=\delta_{ik}\delta_{jr})$, identifying
the two-dimensional basis at each site, labelled by $\{i,j\}$,
with $|0\rangle_{i,j}\;\; (c({i,j})|0\rangle_{i,j}=0)\;$ and
$|1\rangle_{i,j}=c^+({i,j})|0\rangle_{i,j}$.

Then it is not hard to construct fermionic representation of the
$R$ operator. Note, that the $R$-operator defined in the previous
section "permutes" the arrangement of the spaces (as it is a
"check" $R$ matrix), $R:|\alpha'\rangle_1 \; |\beta'\rangle_2
\Rightarrow |\beta\rangle_2\; |\alpha\rangle_1$, so for graded
spaces we have
\bea \label{rcf}{\cal R}=R_{\alpha\;
\beta}^{\beta'\alpha'}|\beta\rangle_2|\alpha\rangle_1\;{
}_2\langle \beta'|{ }_1\langle \alpha'|\\\nn=R_{\alpha\;
\beta}^{\beta'\alpha'}|\beta\rangle_2\langle
\beta'|\;|\alpha\rangle_1\langle
\alpha'|(-1)^{p(\beta')p(\alpha)}, \ena
 where $p(\alpha)=\alpha$ is the
parity of the space $|\alpha\rangle$. Operators
$|\alpha\rangle_i\langle \alpha'|$ act on the Fock space with the
basis $\{|0\rangle_i,\;|1\rangle_i\}$, as
$$
|0\rangle_i \langle 0|=1-c^+_i c_i,\; |1\rangle_i\langle 1|=c^+_i
c_i,\;|0\rangle_i \langle 1|=c_i,\; |1\rangle_i\langle 0|=c^+_i.
$$
This means, that in terms of two scalar fermions $\{c^+_i,\;
c_j\}=\delta_{ij},\quad \{c_i,\; c_j\}=\{c^+_i,\; c^+_j\}=0,\;
i,j=1,2$, the $R$ operator in the zero-field limit, $B=0$, reads
\bea \label{R3} \nn &\mathcal{R}(c^+_1,c_1;c^+_2,c_2)=
R_{00}^{00}+R_{01}^{01}c^+_1 c_2+R_{10}^{10}c^+_2 c_1&\\\nn
&+(R_{10}^{01}- R_{00}^{00})c^+_1c_1 +
(R_{01}^{10}-R_{00}^{00})c^+_2 c_2 +
R_{00}^{11}c^+_2c^+_1&\\\label{R fer}&+ R_{11}^{00} c_2 c_1
+(R_{00}^{00}-R_{10}^{01}-R_{01}^{10}-R_{11}^{11})c^+_1 c_1 c^+_2
c_2,\;\;\;&
\ena
 where $R_{ij}^{kr}$ are the matrix elements of the $R$ matrix.

 \paragraph*{\bf{1. The fermionic representation (\ref{R3})
 of  the Ising model's $R$ matrix}}
 (\ref{rr}) has the following matrix elements:
\bea
R_{00}^{00}&=&2(\cosh{[2J_1]}\cosh{[2J_2]}+1),\nn\\\label{elements}
 R_{11}^{11}&=&2(\cosh{[2J_1]}\cosh{[2J_2]}-1),\\\nn
 R_{01}^{01}&=&R_{10}^{10}=2\sinh{[2J_1]}\cosh{[2J_2]},\\\nn
 R_{10}^{01}&=&R_{01}^{10}=2\sinh{[2J_1]}\sinh{[2J_2]},\\\nn
 R_{00}^{11}&=&R_{11}^{00}=2\cosh{[2J_1]}\sinh{[2J_2]}.
\ena
In the following we shall operate in the coherent-state basis. For
that purpose we need to represent ${\cal R}$ operator in the
normal ordered form. For {\em zero magnetic field} the operator
${\cal R}$ can be expressed as an exponent of a quadratic form (a
consequence of property (\ref{free})),
\bea \label{rf}&&
\mathcal{R}=R_{00}^{00}:\exp{\mathcal{A}(c^+_1,c^+_2,c_1,c_2)}:
\;, \ena
where
\bea\nn &\mathcal{A}(c^+_1,c^+_2,c_1,c_2)=(c-1)(c_1^+ c_1+c_2^+
c_2)&\\&+b(c_1^+ c_2+c_2^+ c_1)+d(c_2^+ c_1^+ + c_2 c_1),& \label{rf1}\\
&a=\frac{R^{00}_{00}}{2},\;b=R_{01}^{01}/R_{00}^{00},\;&\nn\\
&c=R_{01}^{10}/R_{00}^{00},\;d=R_{00}^{11}/R_{00}^{00}.&
\label{abc} \ena

For a {\em finite magnetic field} $B$ the fermionic realization
${\cal R}_B$ of the $R_B$ operator (\ref{Rb}) can be obtained in
the same way, substituting the corresponding matrix elements in
formulas (\ref{rcf}). Then the normal ordered form of the ${\cal
R}_B$ can be represented as follows:
\bea \label{rH} &\mathcal{\cal R}_B=r_B
:e^{\mathcal{A}_B(c^+_1,c^+_2,c_1,c_2)}:,&\\\nn
&\mathcal{A}_B(c^+_1,c^+_2,c_1,c_2)=c_B(c_1^+ c_1+c_2^+
c_2)+b_B(c_1^+ c_2+c_2^+ c_1)\\\nn &+d_B(c_2^+ c_1^+ + c_2
c_1)+h_B(c_1+c_2+c_1^+ +c_2^+)&
\\&+(\eta_{1}c_1+\eta_{2}c_1^+)c_2^+
c_2+(\eta_{2}c_2+\eta_{1}c_2^+)c_1^+ c_1,& \label{awithh}\ena
with
\bea\nn
 r_B&=&2\cosh{[B]} + 2\cosh{[2J_1]}\cosh{[2J_2]}\\\nn
 &+& e^{2(J_1 + J_2)}(\sinh{[B]})^2,\\\nn
 c_B&=&-\frac{2}{r_B}\left(\cosh{[B]} + \cosh{[2(J_1 - J_2)]}\right),\\\nn
 b_B&=&\frac{1}{2 r_B}\left(e^{2(J_1 + J_2)}(\sinh{[B]})^2+2\cosh{[2J_2]}\sinh{[2J_1]}\right)
 ,\\\nn
 d_B&=&\frac{1}{2r_B}\left(e^{2(J_1 + J_2)}(\sinh{[B]})^2+2\cosh{[2J_1]}\sinh{[2J_2]}\right),\\\nn
 h_B&=&\frac{1}{r_B}\left(\sinh{[B]} + \sinh{[2 B]}e^{2(J_1+J_2)}\right),\\\nn
 \eta_1&=&\frac{4}{r_B^2}\left(e^{2J_1}\cosh{[B]} + \cosh{[2J_2]}\right)\sinh{[B]}
 \sinh{[2J_1]},\\
 \eta_2&=&-\frac{4}{r_B^2} \sinh{[B]}\sinh{[2J_1]}\sinh{[2J_2]}.
 \ena
\paragraph*{2. General $R$ matrix and XYZ model.} The fermionic representation (\ref{R3})
is justified as well for arbitrary $R$ matrix, which has form
\bea R=\left(\ba{cccc}R_{00}^{00}&0&0&R_{00}^{11}\\
0&R_{01}^{01}&R_{01}^{10}&0\\
0&R_{10}^{01}&R_{10}^{10}&0\\
R_{11}^{00}&0&0&R_{11}^{11}\ea\right).\label{rg}\ena
Now the $\mathcal{A}(c^+_1,c^+_2,c_1,c_2)$ in the normal ordered
form (\ref{rf}) has a quartic term also,
\bea\label{rfg}&\mathcal{A}(c^+_1,c^+_2,c_1,c_2)=(c-1)c_1^+
c_1+(c'-1)c_2^+ c_2\;\;\;&\\\nn&+b\;c_1^+ c_2+b' c_2^+ c_1
+d\;c_2^+ c_1^+ + d' c_2 c_1+\Delta\; c_1^+ c_1 c_2^+
c_2,&\\\label{abcDelta} &a=\frac{R^{00}_{00}}{2},\;
b=\frac{R_{01}^{01}}{R_{00}^{00}},\;
b'=\frac{R_{10}^{10}}{R_{00}^{00}},\;
c=\frac{R_{10}^{01}}{R_{00}^{00}},\;
c'=\frac{R_{01}^{10}}{R_{00}^{00}},\;\;\;\;\;\;&\\
&d=\frac{R_{11}^{00}}{R_{00}^{00}},\; d'=\frac{R_{00}^{
11}}{R_{00}^{00}},\;
\Delta=\frac{R_{01}^{01}R_{10}^{10}+R_{11}^{00}R_{00}^{
11}-R_{10}^{01}R_{01}^{10}-R_{00}^{00}R_{11}^{11}}{{R_{00}^{00}}^2}.&
\nn \ena
For the $XYZ$ model's general $R$ matrix, given in Eq.
(\ref{rxyz}), the $\Delta$ parameter writes as
\bea
\Delta=2\frac{\mbox{sn}[i\frac{\lambda+u}{2},\;k]\mbox{cn}[i\lambda,\;k]\mbox{dn}[i\lambda,\;k]
} {\mbox{sn}[i\frac{\lambda-u}{2},\;k]
(\mbox{sn}[i\lambda,\;k])^2}.
 \ena
The limit $\Delta=0$ corresponds to the free-fermionic XY model.
It fulfills, when
\be\mbox{cn}[i\lambda,\;k]\mbox{dn}[i\lambda,\;k]=0.\label{cno}\ee
 Possible
solutions are $\mbox{cn}[I,\;k]=0$ and $\mbox{dn}[I+iI'\;k]=0$.
Here $I,\;I'$ are the half-periods of the elliptic functions. Note
that the Ising limit derived in the Ref. \cite{Baxter} corresponds
to the values
\bea\lambda=I'/2,\;\;\;\;\;
\Delta=2\frac{k^{1/2}(1+k)\mbox{sn}[i\frac{\lambda+u}{2},\;k] }
{\mbox{sn}[i\frac{\lambda-u}{2},\;k]}.\label{cis} \ena
%

{\centering
\section{Quantum Field Theory Representation on the Lattice: with general $R$-matrix
and 2DIM}}

In this section we will introduce fermionic fields
$\bar{\psi}(i,j)$ and $\psi(i,j)$, corresponding to the
coherent-states of scalar fermions $c^+(i,j)$ and $c(i,j)$. By
definition, coherent-states are the eigenstates of annihilation
operators of scalar fermions. For a set of the scalar fermions
$c_i$ and $c_i^+$, they are defined by the following relations
\bea \label{psi} c_i|\psi_i\rangle=\psi_i|\psi_i\rangle,\quad
\langle \bar{\psi}_i|c^+_i=\langle \bar{\psi}_i|\bar{\psi}_i. \ena
Because of the Fermi statistics, namely, anticommutation
relations, these eigenvalues are Grassmann variables denoted in
Eq. (\ref{psi}) by $\psi_i$ and $\bar{\psi}_i$. They fulfill ({i})
orthonormality and ({ii}) completeness relations:
\bea
\langle\bar{\psi}_i|\psi_j\rangle=\delta_{ij}e^{\bar{\psi}_i\psi_i},\quad
\int d\bar{\psi}_id\psi_i
e^{-\bar{\psi}_i\psi_i}|\psi_i\rangle\langle \bar{\psi}_i| =
I.\;\;\; \ena
The {\em kernel} of any normal ordered operator $K(\{c^+_i,c_j\})$
in terms of coherent-states can be obtained simply by replacing
creation-annihilation operators $c_i,\; c^+_i$ by their
eigenvalues and multiplying by  $e^{\sum_i \bar{\psi}_i\psi_i}$,
\bea \mathcal{K}(\{\bar{\psi}_i,\psi_j\})\equiv\langle \prod
\bar{\psi}_i|K(\{c^+_i,c_j\})|\prod \psi_j\rangle=\\\nn e^{\sum_i
\bar{\psi}_i\psi_i} K(\{\bar{\psi}_i,\psi_j\}). \ena
The trace of the operator $K(\{c^+_i,c_j\})$ in coherent-states is
an integral over the Grassmann variables,
\bea \nn\mathrm{tr} K(\{c^+_i,c_j\})= \int D\psi D\bar{\psi}
e^{\sum_i\bar{\psi}_i\psi_i}
\mathcal{K}(\{\bar{\psi}_i,\psi_j\}),\\ \quad D\psi
D\bar{\psi}=\prod_i d\psi_i d\bar{\psi}_i.\label{int}
 \ena

In order to obtain the
form of the partition function $Z$ [Eq. (\ref{z})] in the basis of
coherent-states, let us at each circled vertex of the lattice,
between the $R(i,j)$ operators (see Fig. \ref{fig2}), insert the
following identity operators
$$I=\int d\bar{\psi}(i,j)d\psi(i,j)
e^{-\bar{\psi}(i,j)\psi(i,j)}|\psi(i,j)\rangle\langle
\bar{\psi}(i,j)|.$$

With the properties of coherent-states represented above, we can
easily calculate the matrix elements of the ${\cal R}$-operator
(\ref{R3}), using the normal ordered form representations
(\ref{rf}) and (\ref{rfg}), in terms of Grassmann fields,
%
\bea \label{Rcoh} &\langle \bar{\psi}(2i\!+\!1,2j\!+\!1)|\langle
\bar{\psi}(2i\!+\!2,2j)|\qquad&\\\nn
&\qquad\mathcal{R}(i,j)|\psi(2i,2j)\rangle|\psi(2i\!+\!1,2j\!-\!1)\rangle&\\\nn
&=R_{00}^{00}\exp\Bigl\{\mathcal{A}\Bigl[\bar{\psi}(2i\!+\!2,2j),\bar{\psi}(2i\!+\!1,2j\!+\!1),&
\\\nn&\qquad\psi(2i,2j),\psi(2i\!+\!1,2j\!-\!1)\Bigr]&\\\nn
&\!+\!
\bar{\psi}(2i\!+\!2,2j)\psi(2i,2j)\!+\!\bar{\psi}(2i\!+\!1,2j\!+\!1)\psi(2i\!+\!1,2j\!-\!1)\Bigr\}.&
\ena
%
Then the partition function $Z$ for large $N$ can be written as a
path integral, \newline
%
\bea &Z\!=\!\mathrm{tr}
\prod_{j=1}^{N}\prod_{i=N\!-\!1}^{0}\mathcal{R}(i,j)\!=
(R_{00}^{00})^{N^2}\!\int\! D\bar{\psi}D\psi
\;e^{\!-\!A\bigl(\bar{\psi},\;\psi\bigr)}, &\nn\\\label{za}\ena
with action $A(\bar{\psi},\psi)$: \bea \nn
 & \!-\!A(\bar{\psi},\psi)=\sum_{i,j}\{
b\;\bar{\psi}(2i\!+\!2,2j)\psi(2i\!+\!1,2j\!-\!1)&
\\\nn&\!+\!b' \bar{\psi}(2i\!+\!1,2j\!+\!1)\psi(2i,2j)
\!+\!c\;\bar{\psi}(2i\!+\!2,2j)\psi(2i,2j)&
\\\nn&\!+\!c'\;\bar{\psi}(2i\!+\!1,2j\!+\!1)\psi(2i\!+\!1,2j\!-\!1))
&\!\!\!\\\nn&
\!+\!d\;\bar{\psi}(2i\!+\!2,2j)\bar{\psi}(2i\!+\!1,2j\!+\!1)\!+\!d'\;\psi(2i,2j)\psi(2i\!+\!1,2j\!-\!1)&\\\nn&
\!+\!\Delta\;\bar{\psi}(2i\!+\!2,2j)\psi(2i\!+\!1,2j\!-\!1)\bar{\psi}(2i\!+\!1,2j\!+\!1)\psi(2i,2j)&\\\nn&
\!-\!\bar{\psi}(2i,2j)\psi(2i,2j)\!-\!\bar{\psi}(2i\!+\!1,2j\!+\!1)\psi(2i\!+\!1,2j\!+\!1)\}&\\&
\nn\!+\!\sum_{j}\bar{\psi}(2N,2j)\psi(0,2j)\!+\!\sum_{i}\bar{\psi}(2i\!+\!1,2N\!+\!1)\psi(2i\!+\!1,1).&
\\\label{AA} \ena
%
In sum (\ref{AA}) the last two terms come from the trace.

 So, the partition function of a model, defined by $R$ matrix
 (\ref{rg}), has fermionic
path-integral representation with local action (\ref{AA}) on the
two-dimensional lattice.

 As it is apparent, the 2DIM has fermionic representation with local
quadratic action (see Eqs. (\ref{rf1}) and (\ref{elements})). It
is true also for the partition function of the $XY$ model, defined
with the Eqs. (\ref{cno}), as well, since the Gaussian quadratic
form is a consequence of the "free-fermionic" property, given by
Eq. (\ref{free}), and the formula of the coefficient at quartic
term (\ref{abcDelta}). On the other hand, the Ising limit of the
eight-vertex ($XYZ$) model does not correspond to a quadratic
action, as it is followed from the Eqs. (\ref{cis}). However it is
well known that the two limits of the $XYZ$ model - Ising limit
($XZ$) and free-fermionic limit ($XY$) are equivalent and can be
brought one to another by redefinition of the model parameters.

\vspace{0,5cm}

\subsection{Path integral representation of partition function
 for the case of finite magnetic field}

For construction of partition function of 2DIM in a non-zero
magnetic field we make use of fermionic  expression (\ref{rH}). In
order to take into account the graded character of the
$\mathcal{R}(B)$-operators (recall, that they have no definite
parity), we are led to include non local operators,
 \bea \label{zh}
Z(B)\!=\!\mathrm{tr}\prod_{j=1}^{N}
\prod_{i=\!N\!-\!1}^{0}\!\{\mathcal{R}_B^{(even)}(i,j)\!+\!
\mathcal{R}_B^{(odd)}(i,j)J(i,j) \}.\;\;\;\;\;\ena
 In Eq. (\ref{zh}) operators
$\mathcal{R}_B^{(even)}$ and $\mathcal{R}_B^{(odd)}$ represent the
parts of the $\mathcal{R}(B)$-operator (\ref{rH}) that have even
and odd gradings
 (or their series expansions consist of even/odd powers
of $B$ or fermionic operators), correspondingly. Operator
$J(i,j)=\prod(1-2n)$ is the Jordan-Wigner non-local operator (see
Eq. (\ref{saJ}) and the Appendix).

 Let us introduce formal definitions
\bea\nn
&&\mathcal{R'}_B^{(even)/(odd)}(i,j)=\mathcal{R}_B^{(even)/(odd)}(i,j)[1\!-\!2n({\small{2i,2j}})],\\
&&\mathcal{R''}_B^{(even)/(odd)}(i,j)=\nn\\
&&\qquad \qquad [1\!-\!2n(\small{2i\!+\!1,2j\!+\!1})]\mathcal{R}_B^{(even)/(odd)}(i,j).\qquad\qquad\ena
Then we can expand the product in  Eq. (\ref{zh}) and rewrite it
as
 \bea \label{zh1}
Z(B)&=&\mathrm{tr}\sum_{\mathcal{C}_{\{k,r\}}}\prod_{i,j}
\mathcal{\mathbb{R}}^{\mathcal{C}_{kr}}_B(i,j),\ena
where the sum goes over all lattice sites denoted by
$\mathcal{C}_{\{k,r\}}$. Operator
$\mathcal{\mathbb{R}}^{\mathcal{C}_{kr}}_B(i,j)$ is attached to
the square $(i,j)$ (see Fig. \ref{fig2}). It is equal to
$\mathcal{R}_B^{(odd)}(i,j)$ or ${{\mathcal{R}^{'/
''}}}_B^{(odd)}(i,j)$, if $\{i,j\}={\{k,r\}}$, and is equal to
$\mathcal{R}_B^{(even)}(i,j)$ or
${{\mathcal{R}}^{'/''}}_B^{(even)}(i,j)$ otherwise.

 Each summand
in Eq. (\ref{zh1}) can be written in the basis of coherent-states
in the same way as it was done in previous subsection. Finally we
find
\begin{widetext}
\bea\nn Z(B)&=&\int D\bar{\psi}D\psi
e^{-\sum_{i,j}\bar{\psi}(i,j)\psi(i,j)
\!+\!\sum_{j}\bar{\psi}(2N,2j)\psi(0,2j)\!+\!
\sum_{i}\bar{\psi}(2i\!+\!1,2N\!+\!1)\psi(2i\!+\!1,1)\!+\!\mathcal{I}_B(\bar{\psi},\psi)},\\
\mathcal{I}_B(\bar{\psi},\psi)&=&\ln{\sum_{\mathcal{C}_{\{k,r\}}}\prod_{i,j}
\langle \bar{\psi}(2i\!+\!1,2j\!+\!1)|\langle
\bar{\psi}(2i\!+\!2,2j)|\mathcal{\mathbb{R}}^{\mathcal{C}_{kr}}_B(i,j)
|\psi(2i,2j)\rangle|\psi(2i\!+\!1,2j-1)\rangle}.\nn\\
\label{zi}\ena
\end{widetext}

 Because we operate with local fermionic $R_B$ matrices, form
 (\ref{zi}) of the partition function will be held for the case of
 inhomogeneous magnetic field as well.

{\centering
\subsection{Partition function}}
 For calculation of partition function (\ref{za}) in the "free-fermionic"
 case $\Delta=0$ we need to diagonalize
 the action $A(\bar{\psi},\psi)$.
 Taking into account antiperiodic boundary conditions imposed on the
 Grassmann fields,
\bea \nn&\bar{\psi}(2N,2j)=-\bar{\psi}(0,2j),\;
\psi(2N,2j)=-\psi(0,2j),&\\\nn
&\bar{\psi}(2N\!+\!1,2j\!+\!1)=-\bar{\psi}(1,2j\!+\!1),\qquad&\\\nn&
\qquad\psi(2N\!+\!1,2j\!+\!1)=-\psi(1,2j\!+\!1),&\\\nn
&\bar{\psi}(2i,2N)=-\bar{\psi}(2i,0),\;
\psi(2i,2N)=-\psi(2i,0),&\\\nn
&\bar{\psi}(2i\!+\!1,2N\!+\!1)=-\bar{\psi}(2i\!+\!1,1),\qquad&\\&
\qquad\psi(2i\!+\!1,2N\!+\!1)=_\psi(2i\!+\!1,1),& \ena
we can perform the Fourier transformation with odd momenta
\bea \nn \psi(r,k)=\frac{1}{N}\sum_{n_r,n_k=0}^{N-1}e^{-\frac{i
\pi}{2N}((2n_r+1)r+(2n_k+1)k)}\\\times
\psi_\alpha\left[\frac{\pi}{2N}(2n_r+1),\frac{\pi
}{2N}(2n_k+1)\right].\;\;\;\ena
Here $\alpha=1$ for even coordinates $(r,k)$, and $\alpha=2$ for
odd coordinates. After defining new Grassman fields
${\psi}_3(p,q), {\psi}_4(p,q)$ as
\bea &\!\!\!\psi_1(\pi\!-\!p,\pi\!-\!q)\equiv
\!-\!\bar{\psi}_3(p,q),\;
\psi_2(\pi\!-\!p,\pi\!-\!q)\equiv \!-\!\bar{\psi}_4(p,q),\nn \;\;\;\;\;&\\
&\!\!\!\!\!\!\!\!\!\!\bar{\psi}_1(\pi\!-\!p,\pi\!-\!q)\equiv
\psi_3(p,q),\; \bar{\psi}_2(\pi\!-\!p,\pi\!-\!q)\equiv
\psi_4(p,q),\;\;\;\;\;\label{rep}& \ena
we will come to the following simple form for the action $A$
(\ref{AA}) in the momentum space
\bea \label{EQ}
-A(\bar{\psi},\psi)=\sum_{p,q}\sum_{k,r}\mathrm{A}_{kr}(p,q)\bar{\psi}_k(p,q)\psi_r(p,q).
\ena
In Eq. (\ref{EQ}) we have introduced the notations
\be\mathrm{A}(p,q)\!=\! \left(\ba{cccc}
c \;e^{i2p}\!-\!1&b \;e^{i(p+q)}&0&\!-\!d \;e^{i(p\!-\!q)}\\
b' \;e^{i(p+q)}& c'\; e^{2iq}\!-\!1&d \;e^{\!-\!i(p\!-\!q)}&0\\
0&d' \;e^{\!-\!i(p\!-\!q)}&c\; e^{\!-\!2ip}\!-\!1&b' \;e^{\!-\!i(p+q)}\\
\!-\!d' \;e^{i(p\!-\!q)}&0&b \;e^{\!-\!i(p+q)}& c'\;
e^{\!-\!2iq}\!-\!1 \ea\right),\label{A} \ee
and
\bea \nn  p=\frac{\pi}{2N}(2n_p+1),\;\; q=\frac{\pi}{2N}(2n_q+1),\\
n_p=1,...,N/2-1,\;n_q=1,...,N-1. \ena
Then the partition function acquires form of a product of
determinants
\be
Z=(R_{00}^{00})^{N^2}\prod_{n_p,n_q=0}^{\frac{N}{2}-1,N-1}\mathrm{Det}
\left[{\mathrm{A}
\left(\pi\frac{(2n_p+1)}{2N},\pi\frac{(2n_q+1)}{2N}\right)}\right].
\label{det}\ee
 The determinants are found as
\bea \nn
\mathrm{Det}[{\mathrm{A}(p,q)}]=\Big\{1+2\left(\frac{R_{01}^{10}}{R_{00}^{00}}\right)^2+
\left(\frac{R_{11}^{11}}{R_{00}^{00}}\right)^2\Big.\\\nn -
2\frac{R_{01}^{10}(R_{00}^{00}-R_{11}^{11})}{(R_{00}^{00})^2}
(\cos[2p]+\cos[2q])
\nn\\\nn
+4\left(\frac{R_{01}^{10}}{R_{00}^{00}}\right)^2 \cos[2p]\cos[2q]
-2\left(\frac{R_{01}^{01}}{R_{00}^{00}}\right)^2\cos[2(p+q)]\\\label{aDET}
\Big.- 2\left(\frac{R_{00}^{11}}{R_{00}^{00}}\right)^2
\cos[2(p-q)]\Big\}.\;\;\;\;\;\;\;\;\ena
Here we have assumed that $R_{01}^{10}=R_{10}^{01}$ and
$R_{01}^{01}=R_{10}^{10}$.

\paragraph*{Ising model.} Substituting the matrix elements given in
Eqs. (\ref{elements}), we are arriving at
\bea\nn
&\!\!\!\!\!\!\!\!\!\!\!\!\mathrm{Det}[{\mathrm{A}(p,q)}]\{1+\cosh[2J_1]\cosh[2J_2]\}^2
=&\\\nn &2\Big\{1+(\cosh[2J_1]\cosh[2J_2])^2+
(\sinh[2J_1]\sinh[2J_2])^2\Big.&\nn\\\nn& -
2\sinh[2J_1]\sinh[2J_2](\cos[2p]+\cos[2q])&\\\nn&-(\sinh[2J_1])^2
\cos[2(p+q)] -\Big.(\sinh[2J_2])^2
\cos[2(p-q)]\Big\}.&\\\label{DET} \ena
In the limit $p\rightarrow 0,\; q\rightarrow 0$, it takes place
$\mathrm{Det}[{\mathrm{A}(p,q)}]\rightarrow
(\mathrm{Det}[{\mathrm{A}_{0}}])^2$, with
\be \label{DDET}
\mathrm{Det}[{\mathrm{A}_{0}}]=2\left(\frac{1-\sinh[2J_1]\sinh[2J_2]}{1+\cosh[2J_1]\cosh[2J_2]}\right).
\ee
Equations (\ref{DET}) and (\ref{DDET}) suggest that $
\mathrm{Det}[{\mathrm{A}(p,q)}]>0 $ everywhere if $\{p,q\} \neq
\{0,0\}$, and only exactly at $\{p,q\}=\{0,0\}$ and
$\{1-\sinh[2J_{1c}]\sinh[2J_{2c}]=0\}$ we have
$\mathrm{Det}[{\mathrm{A}_{0}}]=0$, corresponding to the point of
the second order phase transition.

\paragraph*{XY model.} Free-fermionic limit of the
eight-vertex model corresponds to relation (\ref{cno}). The
solutions of that relations are (to within the periods of the
elliptic functions)
\be i\lambda=I\;\;\mbox{or}\;\;\;i\lambda=I+iI'. \label{ii'}\ee
When $i\lambda=I$, then the expression in Eq. (\ref{aDET}) writes
as
\bea \label{xy} &\mathrm{Det}[{\mathrm{A}_{xy}(p,q)}] =&
\\\nn &2\Biggl\{1+\left(\frac{\mbox{sn}[i\;u',\;k]\mbox{dn}[i\;
u,\;k]}{\mbox{cn}[i\;u',\;k]}\right)^2(1+2 \cos[2p]\cos[2q])
\Big.& \nn\\
&-\Big.
\frac{\mbox{dn}[i\;u',\;k]^2}{\mbox{cn}[i\;u',\;k]^2}\cos[2(p+q)]
-\left(k\;\mbox{sn}[i\;u',\;k]\right)^2 \cos[2(p-q)]\Biggr\}.&\nn
\ena
Here there is redefinition of the parameter $u$ of the Eq.
(\ref{rxyz}), $u'=\frac{\lambda+u}{2}$. When $k=0$ $XY$ model goes
to the $XX$ model,
\bea \label{xx} \mathrm{Det}[{\mathrm{A}_{xx}(p,q)}]
=\frac{2}{(\cos[u'])^2}\qquad \qquad \qquad \\\nn \times
\Big\{1+2\left(\sin[u']\right)^2\cos[2p]\cos[2q]-
\cos[2(p+q)]\Big\}. \ena
As we can see this expression goes to the value $0$, when
$p=-q=\frac{\pi}{4}$, for all the values of parameter $u'$, which
is a hint of the known fact \cite{Baxter}, that the region $-1\leq
\Delta\leq 1$ corresponds to the critical line of the $XYZ$ model.

 {\centering
\subsection{Continuum limit: IM}}

At the critical line, which is described by the parameters
$J_{1c}$ and $J_{2c}$, correlation length of the system goes to
infinity. All the relevant distances become large at criticality
and it is natural at that limit to
 be interested in large
distances compared to the lattice constant. It is well known that
in the continuum limit, at the point of the second order phase
transition,
$2$DIM is described by free massless Majorana fermions. Below, for
achieving it, we are going to expand the action near the point
$J_{1}=J_{2}=J_c$ (considering for simplicity homogeneous case)
with small values of momenta $p,\;q$.

Diagonalization of matrix (\ref{A}) brings the action to the form
\bea
-A(\bar{\psi},\psi)=\sum_{p,q}\textit{E}_{k}(p,q)\bar{\psi'}_k(p,q){\psi'}_k(p,q),
\ena
with the eigenvalues $\textit{E}_{k}(p,q)$ of $\mathrm{A}(p,q)$
being
%
%
%
\bea 
&\textit{E}_{1}(p,q)\!=\!\mathbf{e}_1\!-\!\mathbf{e}_2\!-\!\mathbf{e}_3,\;
\textit{E}_{2}(p,q)\!=\!\mathbf{e}_1\!-\!\mathbf{e}_2+\mathbf{e}_3,&\nn\\
&\textit{E}_{3}(p,q)\!=\!\mathbf{e}_1+\mathbf{e}_2\!-\!\mathbf{e}_4,\;
\textit{E}_{4}(p,q)\!=\!\mathbf{e}_1+\mathbf{e}_2+\mathbf{e}_4,&
\nn\\ \label{EIGEN}\ena
\bea
&\mathbf{e}_1\!=\!\frac{c}{2}(\cos{[2p]}\!+\!\cos{[2q]})\!-\!1,&\nn\\
&\mathbf{e}_2\!=\!\left(\!
\frac{c^2}{4}(\cos{[2p]}\!\!-\!\!\cos{[2q]})^2\!-\! (d
\sin{[p\!\!-\!\!q]})^2\!-\! (b \sin{[p\!+\!q]})^2\!\right)
^{\frac{1}{2}}\!\!\!,&\nn\\
&\mathbf{e}_3\!=\!\left(\frac{c^2}{4}(\cos{[4p]}\!+\!\cos{[4q]}\!\!-\!\!2)
+ (d \cos{[p\!\!-\!\!q]})^2
\right.\qquad\quad\qquad\;\;\;\;\;\;&\nn\\&\left.+ (b
\cos{[p\!+\!q]})^2
 \!-\! c(\cos{[2p]}\!+\!\cos{[2q]})\mathbf{e}_2 \right) ^{\frac{1}{2}},&\nn\\&
 \mathbf{e}_4\!=\!\left(\frac{c^2}{4}(\cos{[4p]}\!+\!\cos{[4q]}\!\!-\!\!2)+
(d
\cos{[p\!\!-\!\!q]})^2\right.\qquad\quad\qquad\;\;\;\;\;\;&\nn\\&+
\left.(b\cos{[p\!+\!q]})^2
 + c(\cos{[2p]}\!+\!\cos{[2q]})\mathbf{e}_2\right)
 ^{\frac{1}{2}}.&\nn
 \ena
We see from Eq. (\ref{EIGEN}) that at the critical value of
coupling $J$, $J=J_c$, and at the momenta  $\{p,q\}=\{0,0\}$ (or
$\{p,q\}=\{0,\pi\}$) two eigenvalues $\textit{E}_{2}(p,q)$ and
$\textit{E}_{4}(p,q)$ become $0$, whereas the remaining two
eigenvalues $\textit{E}_{1}(p,q)$ and $\textit{E}_{3}(p,q)$ take
the value $-\frac{4}{3}$. As we have mentioned, taking the
continuum limit at the point of second order phase transition is
justified, as the lattice constant can be neglected compared to
the correlation length, and the latter is proportional to the
inverse of mass. Thus, we expand  the action for the massless
fermions $\bar{\psi'}_k(p,q),\;{\psi'}_k(p,q)$, $k=2,4$,  at the
critical point. Expansion of the eigenvalues gives
\bea\textit{E}_{2}(p,q)=\sqrt{2}(J-J_c)-\sqrt{-p^2-q^2},\\\nn
\textit{E}_{4}(p,q)=\sqrt{2}(J-J_c)+\sqrt{-p^2-q^2}.\ena
After a linear transformation of the field variables
$\psi'_2(p,q)$ and ${\psi'}_4(p,q)$, the sum
\bea\left(\sqrt{2}(J-J_c)-\sqrt{-p^2-q^2}\right)\bar{\psi'}_2(p,q){\psi'}_2(p,q)\nn
\\+
\left(\sqrt{2}(J-J_c)+\sqrt{-p^2-q^2}\right)\bar{\psi'}_4(p,q){\psi'}_4(p,q)\nn\ena
takes the form
\bea &\left(\ba{c}\bar{\psi}_{+}(p,q),
\bar{\psi}_{-}(p,q)\ea\right)\left(\ba{cc}m&iq-p\\
iq+p&m
\ea\right)\left(\ba{cc}\psi_{+}(p,q)\\
\psi_{-}(p,q) \ea\right), \;\;\;\;\;\;&\\\nn & m=\sqrt{2}(J-J_c).
& \ena
Continuum action of 2DIM can be conveniently written upon
introducing two-dimensional gamma matrices $\gamma^0=\sigma_1$ and
$\gamma^1=i\sigma_2$,  as
\bea -A(\bar{\psi},\psi)=\int
\bar{\psi}(p,q)\left(m-i\gamma^{\mu}p_{\mu}\right)\psi(p,q).
\label{contin}\ena
Here $ \psi(p,q)\!=\!\left(\!\ba{cc}\psi_{+}(p,q)\\ \psi_{-}(p,q)
\ea\!\right), \;
\bar{\psi}(p,q)\!=\!\left(\!\ba{cc}\bar{\psi}_{+}(p,q)&
\bar{\psi}_{-}(p,q) \ea\!\right)$, $p_0=i q$ and $p_1=p$.

\vspace{0.5 cm}

{\centering
\subsection{Thermal capacity}}
Determinant representation of partition function (\ref{det}) leads
to the following expression for free energy, $F=-(T \ln{Z})/N^2$,
per site:
\bea \nn
&F=\!- T/N^2\sum_{p,q}\ln{2\{1\!+\!(\cosh[2J_1]\cosh[2J_2])^2}&\nn\\
&\!+\!(\sinh[2J_1]\sinh[2J_2])^2\!&\\\nn
&-2\sinh[2J_1]\sinh[2J_2](\cos[2p]\!+\!\cos[2q])&\nn\\
\nn &\!-(\sinh[2J_1])^2
\cos[2(p\!+\!q)]\!-\!(\sinh[2J_2])^2\cos[2(p\!-\!q)]\}.&
\label{FF} \ena
The thermal capacity is related to the second derivative of the
free energy with respect to the temperature as follows:
\bea \label{CCT}
C=-T\frac{\partial^2 F}{\partial T^2}. \ena
In order to obtain  the temperature dependence of the free energy,
one has to replace parameters $(J_1, J_2)$ with $(J_1/T,J_2/T)$.
Then the result for thermal capacity $C$ follows upon performing
this replacement in Eq. (\ref{FF}) and substituting $F$ into Eq.
(\ref{CCT}). The result has a simple form in homogeneous case,
$J_1=J_2=J$, and reads
\bea \label{Csum}  C=4\frac{1}{N^2} \left(\frac{J}{T}\right)^2
\sum_{p,q}^{N,N/2}\qquad\\\nn
 \left\{2\frac{\cosh[4\frac{J}{T}]+
\cosh[8\frac{J}{T}]-4(\cos[p]\cos[q])^2
\cosh[4\frac{J}{T}]}{(\cosh[2\frac{J}{T}])^4-4(\cos[p]\cos[q]\sinh[2\frac{J}{T}])^2}
\right.\\ -\left. \left(\frac{(1+\cosh[4\frac{J}{T}]-4(\cos[p]
\cos[q])^2)\sinh[4\frac{J}{T}]}{(\cosh[2\frac{J}{T}])^4-4
(\cos[p]\cos[q])^2}\right)^2\right\}.\nn \ena
In the thermodynamic limit, $N\rightarrow \infty$, the sum in Eq.
(\ref{Csum}) should
 be replaced by the integral as
\bea \frac{1}{N^2}\sum_{p,\;q}^{N/2,N} \rightarrow
\frac{1}{\pi^2}\int_0^{\pi/2}\int_0^{\pi}dp \; dq. \label{si} \ena
Then, after performing the integration, we obtain
\bea \!\!\!\!\!\!\!\!C= \label{OO} \frac{4}{\pi}\left(\frac{ J}{
T} \; \mb{csch}\Big[4\frac{ J}{T}\Big]\right)^2
\left\{-4\left(\cosh\Big[2\frac{J}{T}\Big]\right)^2\right.\qquad
\\ \nn \times\left(\pi+\Big(1+\cosh\Big[4\frac{
J}{T}\Big]\Big) \mathbf{E}\left[ 4\;
\left(\mb{sech}\Big[2\frac{J}{T}\Big]\tanh\Big[2\frac{J}{T}\Big]\right)^2\right]\right)\\\nn
+\left.\left(15+\cosh\Big[8\frac{J}{T}\Big]\right)
\mathbf{K}\left[4\; \left(\mb{sech}\Big[\frac{2J}{T}\Big]
\tanh\Big[\frac{2J}{T}\Big]\right)^2\right]\right\}. \ena
Here the functions $\mathbf{E}$ and $\mathbf{K}$ are the complete
elliptic integrals of the second and the first kinds. Equation
(\ref{OO}) reproduces the expression for the thermal capacity
obtained from Onsager's solution \cite{O,KH,MV}. The consequence
of the factorization property of the determinants in expression
Eq. (\ref{det}) for the partition function,
\bea \label{p1}
&\mathrm{Det}[{\mathrm{A}(p,q)}](1+\cosh[2J_1]\cosh[2J_2])^2/4
&\\&\nn =(\cosh{[2J_1]}\cosh[2J_2]
\!-\!\cos[p\!+\!q]\sinh[2J_1]&\\&\nn \!-\!\cos[p\!-\!q]
\sinh[2J_2])
 \times (\cosh[2J_1]\cosh[2J_2]&\\&\nn \!+\!\cos[p\!+\!q]\sinh[2J_1]\!+\!
\cos[p\!-\!q]\sinh[2J_2]),&\ena
demonstrates the link to Onsager's solution \cite{O}. Note, that
the first and the second terms in the product on right-hand side
of Eq. (\ref{p1}) differ only by shifts $\pi-\bar{p}$ and
$\pi-\bar{q}$, where $\bar{p}=p+q$ and $\bar{q}=p-q$. Therefore,
expression (\ref{det}) for the partition function can be written
as a product of the first terms in Eq. (\ref{p1}) { only}, where
$\bar{p}=p+q$ and $\bar{q}=p-q$ take values in the interval from
zero to $\pi$.
%
\vspace{0.5cm}
\section{Correlation functions: IM, $B=0$}

Fermionic approach formulated above is very convenient for
calculation of  correlation functions and spontaneous
magnetization. Let us first analyze vacuum expectation value of
the spin variable, $\bar{\sigma}_\alpha $,
\bea \label{aversigma}
\langle\bar{\sigma}_{\alpha}(i,j)\rangle=\frac{1}{ Z }
\sum_{\{\bar{\sigma}\}} \{\bar{\sigma}_\alpha(i,j)\prod_{k,r}
W_{\alpha'\beta'}^{\alpha''\beta''}(k,r)\}. \ena
Here, as in the beginning, $\bar{\sigma}_\alpha(i,j)$ are
classical spin variables attached to the vertex $(i,j)$.

Our recipe for further evaluation is simple. For calculation of
the average of any quantity, say
$\bar{g}(\{\bar{\sigma}_\alpha(i,j)\})$, first we represent it in
the spin operator form (as it was done for the Boltzmann weights
in Sec. I) as a function of Pauli operators,
$g(\{{\sigma}_k(i,j)\})$. Then we determine corresponding
fermionic realization of $g$ in the normal ordered form
$N\left[g^f\bigl(\{c^+(i,j),c(i,j)\}\bigr)\right]$. Average
$\langle\bar{g}(\{\bar{\sigma}_\alpha(i,j)\})\rangle$ then will be
equivalent to the Green's function $\langle
N[g^f(\{\bar{\psi}(i,j),\psi(i,j)\})]\rangle$ in the corresponding
fermionic field theory with local quadratic action (\ref{AA}) on
the lattice.

 The average of a spin variable in the Eq. (\ref{aversigma}) can
be expressed via operator forms of Boltzmann weights (\ref{W1})
and $R$-matrices (\ref{R1}),
\bea \nn &\langle\bar{\sigma}_{\alpha}(i,j)\rangle\!\!
=\!\!\frac{1}{ Z }\; \mb{tr} \prod_{\{k,\;r>j\}}\!\! W(k,r)
\!\prod_{\{k>i\}}\!\! W(k,j)\; \sigma_z(i,j)\!
&\quad\\&\nn\prod_{\{k,\;r\leq j\}}\!\! W(k,r) \!\prod_{\{k\leq
i\}}\!\! W(k,j)&\\& \nn \!\!=\!\! \frac{1}{ Z}\;
 \mb{tr}
\prod_{\{k,\;r>j\}}\!\!R(k,r)\!\prod_{\{k>i\}}\!\!R(k,j)
\;\sigma_1(i,j)\! &\\&\prod_{\{k,\;r\leq j\}}\!\!
R(k,r)\!\prod_{\{k\leq i\}}\!\! R(k,j). & \label{ss} \ena
Here the trace is understood as the composition of $\mathrm{tr}_a$
defined in Eqs. (\ref{tr1}) and (\ref{tr2}):
$\mathrm{tr}=\mathrm{tr}_2 \mathrm{tr}_1$. By taking into account
Jordan-Wigner non-local operator, $J=\prod(1-2n)$ (for details,
see the Appendix), we can represent the single spin operators on
the lattice via fermionic creation-annihilation operators,
\bea
&&\!\!\!\!\!\!\sigma_1(\mb{\small 2\textit{i},
2\emph{j}})\rangle= [c(\mb{\small 2\emph{i},
2\emph{j}})+c^+(\mb{\small 2\emph{i},
2\emph{j}})]J(i,j),
\label{sa}\\
&&\!\!\!\!\!\! \sigma_1(\mb{\small
2\textit{i}+1, 2\emph{j}+1})\rangle= [c(\mb{\small 2\emph{i}+1,
2\emph{j}+1})\nonumber\\
&&\qquad \qquad + c^+(\mb{\small 2\emph{i}+1, 2\emph{j}+1})]
J(i,j),\label{sa1}\\&&\!\!\!\!\!\!
J(i,j)=\prod_{k<i}[1-2n({\mb{\small {2\emph{k}+1, 2\emph{j}+1}}})]\nn\\
&&\qquad\cdot\prod_{r>j}[1-2n(\mb{{\small 0, 2\emph{r}+2}})].\label{saJ} \ena
For a finite lattice the expectation value given by Eq. (\ref{ss})
always acquires value $0$ due to the $\mathcal{Z}_2$ symmetry of
the model. In fermionic approach this is quite apparent, as it
corresponds to an integration of a polynomial over odd Grassmann
variables (see Eqs. (\ref{sa}) and (\ref{sa1})), while the
integration goes by even number of variables, Eq. (\ref{int}). The
case of infinite lattice will be specified in the next section.

\begin{figure}[h]
\unitlength=22pt
\centerline{\includegraphics{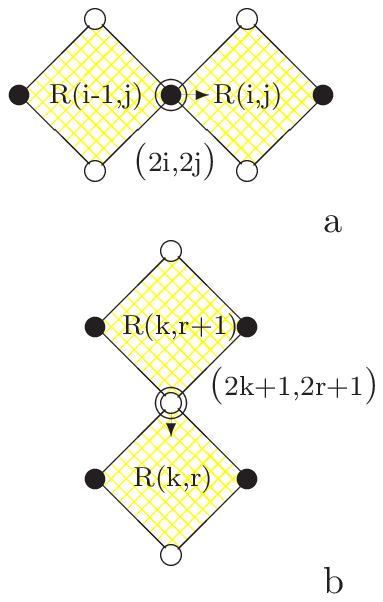}}
\caption{{$\;\;\;$}(a) $\mathcal{R}(i,j)[\;c^+(2i,2j)\pm c(2i,2j)]$\\
{$\;\;\;\;\;\;\;\;$} (b)
 $[\;c^+(2k+1,2r+1)\pm c(2k+1,2r+1)]\mathcal{R}(i,j)$}\label{fig3}
\end{figure}

Now it is convenient to rewrite operators $(1-2n)$  as
\be 1-2n=(c^++c)(c^+-c), \label{1n}\ee
which brings expressions (\ref{sa}) and (\ref{sa1}) to the form $
(c^++c)\left(\prod(c^++c)(c^+-c)\right)$. Then we insert the
resulting formulas of Eqs. (\ref{sa}) and (\ref{sa1}) into Eq.
(\ref{ss}). In the previous section we included fermionic fields
for each $R$ operator \textit{locally} (or for each dashed square
in the lattice on Fig. \ref{fig2}), later represented them in the
normal ordered form and finally switched to the coherent basis. In
order to escape complications in the further calculations, we
shall always attach "even-even" $[c^+(2i,2j)\pm c(2i,2j)]$
fermionic operators to $R(i,j)$ matrix, [Fig. \ref{fig3} a], and
the "odd-odd" fermionic operators $[c^+(2k+1,2r+1)\pm
c(2k+1,2r+1)]$ to $R(k,r)$-matrix [Fig. \ref{fig3} b]. In Fig.
\ref{fig3}
operators $(c^+\pm c)$ are shown by large circles on the vertices.
This choice, which of course will not affect the
 result of the calculation of expectation values, has a simple
 explanation.

 %

Let
 us consider normal ordered forms of the operators
$\mathcal{R}(i,j)[c^+(2i,2j)\pm c(2i,2j)]$ and
 $[c^+(2k+1,2r+1)\pm c(2k+1,2r+1)]\mathcal{R}(i,j)$,
  where $\mathcal{R}$  is the fermionic $R$ operator given by
Eq. (\ref{R fer}). They are particular cases of a general
expression
\bea \label{xi} R_{00}^{00} :[x_1 c_1+x_2c_2+x_3 c_1^+ +x_4 c_2^+]
e^{\mathcal{A}(c_1^+,\;c_2^+,\;c_1,\;c_2)}:, \;\;\;\ena
with different choice of $x_1, x_2, x_3$ and $x_4$ depending on
parameters $b,\; c,\; d$ [see Eq. (\ref{abc})]. $\mathcal{A}$ is
defined by Eq. (\ref{rf1}).

 While for operators
$[c^+(2i,2j)\pm c(2i,2j)]\mathcal{R}(i-1,j)$ and
 $\mathcal{R}(i,j+1)[c^+(2k+1,2r+1)\pm c(2k+1,2r+1)]$,
fermionic normal ordered form belongs to the following general
expression
\bea R_{00}^{00}:[x(c_1+c_1^+ +c_2+ c_2^+) + (x_1 c_1 +x_2
c_1^+)c_2^+ c_2\nn\\+ (x_3 c_2+ x_4 c_2^+)c_1^+ c_1]
e^{\mathcal{A}(c_1^+,\;c_2^+,\;c_1,\;c_2)}:. \label{rc}\ena
  Eqs. (\ref{xi})
and (\ref{rc}) show that in the latter case
we have additional powers of Grassmann fields.

In expressions of the two-point operators
$\sigma_1(i,j)\sigma_1(k,r)$  some of $(1-2n)$ operators coincide
and cancel each other due to Eq.
(\ref{n1}). The remaining operators between two points, $(i,j)$
and $(k,r)$, form a path, which can be deformed using feature
(\ref{n2}). For example, if $i_1>i_2,\;\; j_2>j_1$, we have
\bea\label{cc}
&\sigma_1(2i_2\!+\!1,2j_2\!+\!1)\sigma_1(2i_1,2j_1)&\!\!\!\!\!\!\!\!\!\!\!\!\!\!\!\!\\\nn&
=[c(2i_2\!+\!1,2j_2\!+\!1)\!+\!c^{\!+\!}(2i_2\!+\!1,2j_2\!+\!1)]&\\&\nn
\times
[1-2n(2i_2\!+\!1,2j_2\!+\!1)]\prod_{r=j_1\!+\!1}^{j_2}[1\!-\!2n(2i_2\!+\!2,2r)]&\\&\nn\times
\prod_{k=i_2\!+\!1}^{i_1-1}[1-2n(2k\!+\!1,2j_1\!+\!1)]
[c(2i_1,2j_1)\!+\!c^{\!+\!}(2i_1,2j_1)]&\\&\nn
=[c^{\!+\!}(2i_2\!+\!1,2j_2\!+\!1)\!-\!c(2i_2\!+\!1,2j_2\!+\!1)]&\\&\nn\times
\prod_{r=j_1\!+\!1}^{j_2}[1\!-\!2n(2i_2\!+\!2,2r)]&\\&\nn \times
\prod_{k=i_2\!+\!1}^{i_1-1}[1\!-\!2n(2k\!+\!1,2j_1\!+\!1)]
[c(2i_1,2j_1)\!+\!c^{+}(2i_1,2j_1)].&\ena
%
Making use of expression (\ref{1n}),
we bring the correlation functions
$\langle\sigma_1(i_1,j_1)\sigma_1(i_2,j_2)\rangle$ to the form
\bea\langle[c^+-c]\left[\prod(c^++c)(c^+-c)\right][c^++c]\rangle.\label{cpc}\ena
With the help of the Wick's theorem, we can represent average
(\ref{cpc})
in terms of the Pfaffian form with elements
$$\langle [c^+(i,j)\pm
c(i,j)][c^+(k,r)\pm c(k,r)]\rangle. $$

Let us consider in details the cases, when two spins are arranged
along a direct line on the lattice, in horizontal or in vertical
directions. Since there is a translational invariance in both
directions we shall restrict ourselves by two cases:
$\langle\sigma_1(0,0)\sigma_1(0,2k)\rangle$ and
$\langle\sigma_1(2k+1,1)\sigma_1(1,1)\rangle$. For the vertically
arranged spins we have
\bea\nn & G(k)\equiv\langle\sigma_1(0,2k)\sigma_1(0,0)\rangle
=\langle[c^{\!+\!}(0,2k)\!+\!c(0,2k)]&\\\nn&\times\prod_{r=0}^{k-1}[1\!-\!2n(0,2r)]
[c^{\!+\!}(0,0)\!+\!c(0,0)]\rangle&\\\nn &=\langle
\left[c^{\!+\!}(0,2k)\!+\!c(0,2k)\right]&\\\nn&\times\left(\prod_{r=1}^{k-1}
[c(0,2r)\!-\!c^{\!+\!}(0,2r)][c^{\!+\!}(0,2r)\!+\!c(0,2r)]\right)&\\&
\label{GJ12}\times [c(0,0)-c^{\!+\!}(0,0)]\rangle.&\ena
Wick's rules allow representing the last expression in Eq.
(\ref{GJ12}) as a square root of a determinant, and hence $G(k)$
has the following representation by means of a Gaussian path
integral:
\bea G(k)\!=\!\int \! D\chi\; e^{\frac{1}{2}\sum_{i,j=1}^{2k}
\mathcal{G}_{ij}\;\chi(i)\chi(j)-\sum_{i=0}^{2k-1}
\chi(2i\!+\!1)\chi(2i)}.\;\;\;\;\;\label{gint}\ena
Here $\chi(i)$'s  are Grassmann variables.  Antisymmetric
 matrix elements $\mathcal{G}_{ij}$
are defined as
\bea \nn&& \mathcal{G}_{2i\!+\!1\;
2j\!+\!1}=\langle[c^{\!+\!}(0,2i)\!-\!c(0,2i)][c^{\!+\!}(0,2j)\!-\!c(0,2j)]\rangle,\\\label{G1}
&&\mathcal{G}_{2i\;2j}=\langle[c^{\!+\!}(0,2i)\!+\!c(0,2i)][c^{\!+\!}(0,2j)\!+\!c(0,2j)]\rangle,\;\;\;\;\;\;\;\;\;\;\;\;
\\\nn
&&\mathcal{G}_{2i\;2j\!+\!1}=\langle[c^{\!+\!}(0,2i)\!+\!c(0,2i)][c(0,2i)\!-\!c^{\!+\!}(0,2j)]\rangle.
\ena
 All the expressions in (\ref{G1}) can be easily derived in
the basis of coherent-states (\ref{psi}). As it was stated
earlier, the normal ordered form of the operator
$\mathcal{R}(i,j)[c^+(2i,2j)\pm c(2i,2j)]$  has form (\ref{xi}).
The parameters in this case are given by
 \bea
\{x_1,x_2,x_3,x_4\}=\{\pm1,d,c,b\}.\ena
Let $\{x_1',x_2',x_3',x_4'\}$ and $\{x''_1,x''_2,x''_3,x''_4\}$ be
the parameters corresponding to the operator $(c^+\pm c)$ at
$(0,2i)$ and $(0,2j)$ points, respectively. Then we can rewrite
all the expressions in Eq. (\ref{G1}) as a general function of
these parameters, namely $\mathcal{G}(i,j,\{x'\},\{x''\})$, which
has the following integral form in coherent-state basis:
\bea\label{Ga}
&\mathcal{G}(i,j,\{x'\},\{x''\})=\frac{(R_{00}^{00})^{N^2}}{Z}
\int D\bar{\psi}D\psi\;
e^{A(\bar{\psi},\psi)}&\\&\nn\times\left[x'_1
\psi(0,2i)\!+\!x'_2\psi(1,2i\!-\!1\!)+x'_3 \bar{\psi}(2,2i) \!+\!
x'_4 \bar{\psi}(\! 1,2i\!+\!1\!) \right]&\\& \times \left[x''_1
\psi(0,2j)\!+\!x''_2\psi(1,2j\!-\!1\!)+x''_3 \bar{\psi}(2,2j)\!+\!
 x''_4\bar{\psi}(1,2j\!+\!1\!)
 \right],&\nn
\ena
where $A$ and $Z$ are defined in Eqs. (\ref{AA}) and (\ref{za}).

The second sum ({\small$-\sum_{i=0}^{k-1} \chi(2i+1)\chi(2i)$}) in
Eq. (\ref{gint}), and hence the additional unity elements with
$\mathcal{G}_{2i+1\;2i}(-\mathcal{G}_{2i,\;2i+1})$, are
conditioned by the normal ordered version of relation
$1-2n=(c^++c)(c^+-c)$, i.e. $\;
:\!1-2n\!\!:=1+\!:\!(c^++c)(c^+-c)\!:$.

Then straightforward calculations lead to the following expression
for $\mathcal{G}(r, j,\{x'\},\{x''\})$:
\bea \nn\!\!\! &&\mathcal{G}(r,j,\{x'\},\{x''\}) =
1/N^2\sum_{n_1=1,n_2=1}^{N/2,N}\\\nn
 &&\left\{\left(\frac{K_1}{k_1}  x_3' x_4'' \!+\! \frac{k}{K_1} x_1' x_2'' \!-\!
\frac{k}{K_1} x_4' x_3'' \!-\! \frac{K_1}{k_1} x_2'
x_1''\right)(\mathrm{A}^{\!-\!1})_{14}\right.
\\\nn &&\!+\!\left(K_1 x_3'
x_2'' \!-\! \frac{k k_1}{K_1} x_1' x_4'' \!+\! K_1 x_4' x_1''
\!-\!\frac{ k k_1}{K_1} x_2'
x_3''\right)(\mathrm{A}^{\!-\!1})_{12}\\\nn &&
\!+\!\left(\frac{K_1}{k} x_4' x_3'' \!+\! \frac{k_1}{K_1} x_2'
x_1'' \!-\! \frac{k_1}{K_1} x_3' x_4'' \!-\! \frac{K_1}{k} x_1'
x_2''\right) (\mathrm{A}^{\!-\!1})_{23}\\\nn &&\!+\!
\left(\frac{1}{K_1} x_3' x_2'' \!-\! \frac{K_1}{k k_1} x_1' x_4''
\!+\! \frac{1}{K_1} x_4' x_1'' \!-\! \frac{K_1}{k k_1} x_2' x_3''
\right)(\mathrm{A}^{\!-\!1})_{34}\\\nn
&&\!+\!\left(\!K\!-\!\frac{1}{K}\!\right)\Big(\!( x_3' x_3'' \!-\!
x_1' x_1'')(\mathrm{A}^{\!-\!1})_{13}+( x_4' x_4'' \!-\! x_2'
x_2'')(\mathrm{A}^{\!-\!1})_{24} \!\Big) \\\nn&&\!+\! k\left(K
x_3' x_1'' \!-\! \frac{x_1'
x_3''}{K} \right)(\mathrm{A}^{\!-\!1})_{11}
\!+\! k_1\left(K x_4' x_2'' \!-\! \frac{ x_2' x_4''}{K}\right)
(\mathrm{A}^{\!-\!1})_{22}
\\\nn &&\!+\! \frac{1}{k}\left(\frac{x_1'' x_3'}{K}  \!-\!
K x_1' x_3''\right)(\mathrm{A}^{\!-\!1})_{33} \\\nn &&\left. \!+\!
\frac{1}{k_1}\left(\frac{x_4' x_2''}{K} \!-\! K x_2' x_4''\right)
(\mathrm{A}^{\!-\!1})_{44}  \right\}\Big|
_{p=2\pi\frac{2n_1\!+\!1}{2N},q=2\pi\frac{2n_2\!+\!1}{2N}}.\\\label{skz}
\ena
 Here
$\mathrm{A}$ is the $4\times 4$ matrix defined by Eq. (\ref{A}),
and
\bea K= e^{{{i}}(2r p\!+\!2j q)},\quad K_1=e^{{
i}((2r\!+\!1)p\!+\!(2j\!+\!1)q)},\\\nn k=e^{{ i}\; 2 p },\quad
k_1=e^{{ i} \;2 q }, \qquad r,j= 1 \cdots N .\ena

Similar expressions can be obtained for the horizontally arranged
spins too,
\bea
\label{g'}&&G'(k)\equiv\langle\sigma_1(2k\!\!+\!\!1,1)\sigma_1(1,1)\rangle=
\\\nn &&=\langle
[c^{+}(2k\!+\!1,1)\!-\!c(2k\!+\!1,1)]\\\nn &&\times
\prod_{r=1}^{k-1}[1-2n(2r\!+\!1,1)] [c^{+}(1,1)+c(1,1)]\rangle
 ,\ena
which also admits integral representation (\ref{gint}), in this
case with the matrix elements
\bea\nn \mathcal{G}_{2i\!+\!1,2j\!+\!1}\qquad\\\nn
=\langle[c^{\!+\!}(2i\!+\!1,1)\!-\!c(2i\!+\!1,1)][c^{\!+\!}(2j\!+\!1,1)\!-\!c(2j\!+\!1,1)]\rangle,\\\nn
\mathcal{G}_{2i,2j}\qquad\\\nn=\langle[c^{\!+\!}(2i\!+\!1,1)\!+\!c(2i\!+\!1,1)][c^{\!+\!}(2j\!+\!1,1)\!+\!c(2j\!+\!1,1)]\rangle,
\\\nn
\mathcal{G}_{2i,2j\!+\!1}\qquad\\\nn=\langle[c^{\!+\!}(2i\!+\!1,1)\!-\!c(2i\!+\!1,1)][c^{\!+\!}(2j\!+\!1,1)\!+\!c(2j\!+\!1,1)]\rangle.
\\\label{G2}\ena
The elements given by Eq. (\ref{G2}) can also be expressed by the
function  $\mathcal{G}(i,j,\{x'\},\{x''\})$  [Eqs. (\ref{Ga}) and
(\ref{skz})], but here the parameters defined by the normal
ordered form (\ref{xi}) of the operator $[c^+(2i+1,2j+1)\pm
c(2i+1,2j+1)]\mathcal{R}(i,j)$ read
 \bea
\{x_1,x_2,x_3,x_4\}=\{b,c,d,\pm 1\}
.\ena
   The above relations enable us to find correlation functions for all the
statistical models with weights, which can be written in the
matrix form (\ref{rg}), with  condition (\ref{free}), letting
$R_{01}^{10}=R_{10}^{01},\;R_{01}^{01}=R_{10}^{10}$.

\paragraph*{2DIM.} Inserting the parameters $\{x'\},\;\{x''\}$ and the elements
of the inverse matrix $\mathrm{A}^{\!-\!1}$ defined for the 2DIM
(\ref{elements}) into Eq. (\ref{skz}), we find the following
expression for Eq. (\ref{G2})
{\small \bea \nn &&
\mathcal{G}_{2i\;2j}=\mathcal{G}_{2i\!+\!1\;2j\!+\!1}=\frac{\!-\!2}{N^2}\sum_{n_1=1}^{N/2\!-\!1}
\sum_{n_2=1}^{N\!-\!1}\frac{\sin[2(i\!-\!j)p\;]}{a^2
\mathrm{Det}[{A(p,q)}]}\qquad
\\\nn &&\times\Big\{
\cosh{[2J_2]}\sin{[2(p\!-\!q)]}\!-\!\cosh{[2J_1]}\sin{[2(p\!+\!q)]}
\\\nn && \!+\! 2\sin{[2q]}(\cos{[2p]}\!-\!\!2\sinh{[J_1]}\sinh{[J_2]})
\Big\},\\\nn &&
\mathcal{G}_{2k\;2r\!+\!1}=\!-\!\mathcal{G}_{2r\!+\!1\;2k}=\frac{2}{N^2}\sum_{n_1=1}^{N/2\!-\!1}
\sum_{n_2=1}^{N\!-\!1}\frac{1}{a^2 \mathrm{Det}[{A(p,q)}]}\qquad
\\\nn &&\times\Big\{\cos{[2(r\!-\!k)p]}\Big(3
\!+\!\cosh{[2J_1]}\cosh{[2J_2]}\\\nn && \!-\!
4\cosh{[J_1]}\cosh{[J_2]} \!+\! 2\cos{[2 p]} \cos{[2 q]}
\Big.\qquad\\\nn &&\!-\! \cos{[2(p \!+\! q)]}\cosh{[2
J_1]}\!-\!\cos{[2(p \!-\! q)]}\cosh{[2 J_2]}\\\nn && \!-\!
4(\cos{[2 p]} \!+\!\cos{[2 q]})\sinh{[J_1]}\sinh{[J_2]}\Big)
\\ &&\Big. \!+\!
\cos{[2(r\!-\!k\!-\!1)p]}\sinh{[2 J_1]}\sinh{[2
J_2]}\Big\},\label{j12} \ena \bea {p= \frac{\pi}{2N} (2n_1 \!+\!
1),\qquad q= \frac{\pi}{2N} (2n_2 \!+\! 1)}.\nn \ena}
The elements $\mathcal{G}_{ij}$ in Eq. (\ref{G1}) for the vertical
case can be obtained from expressions (\ref{j12}) simply by
interchanging the coupling constants $J_1$ and $J_2$.

 In the homogeneous case $J_1=J_2=J$, we have  $\mathcal{G}_{2i
 \;2j}=0$ and
 $\mathcal{G}_{2i\!+\!1\;2j\!+\!1}=0$,
 and the expression for $G(i)$ [$G'(i)$] simplifies to the determinant
\bea
\label{Gi1}G(i)=\mathrm{Det}\left(\ba{cccc}\bar{\mathcal{G}}_{i
0}&\bar{\mathcal{G}}_{i1}&\cdots
&\bar{\mathcal{G}}_{ii\!-\!1}\\
\bar{\mathcal{G}}_{i\!-\!10}& \bar{\mathcal{G}}_{i\!-\!1 1}& \cdots & \bar{\mathcal{G}}_{i\!-\!1 i\!-\!1}-1\\
\vdots&&&\vdots\\
\bar{\mathcal{G}}_{10}&\bar{\mathcal{G}}_{11}-1& \cdots &
\bar{\mathcal{G}}_{1i\!-\!1}\ea\right)\!,\;\;\;\;\;\;
 \ena \bea
&&\bar{\mathcal{G}}_{k\!\!+\!\!i\;k}\equiv
\bar{\mathcal{G}}_{i}\equiv
\mathcal{G}_{2(k\!+\!i)\;2k\!+\!1}.\ena
Therefore we can rewrite the Gaussian integral representation
(\ref{gint}) in the following way:
\bea \label{Gii} G(i)=\int D\bar{\chi'}D\chi {\; e
\;}^{\sum_{k=1,r=0}^{i,i-1}\bar{\mathcal{G}}_{k-r}\bar{\chi'}_k
\chi_r-\sum_{k=1}^{i-1}\bar{\chi'}_k \chi_k}. \;\;\;\;\;\ena
After the replacement $\bar{\chi'}_k=\bar{\chi}_{k-1}$, Eq.
(\ref{Gii}) reads
 \bea G(i)=\int D\bar{\chi}D\chi
{\; e
\;}^{\sum_{k,r=0}^{i-1}\bar{\mathcal{G}}_{k-r\!+\!1}\bar{\chi}_k
\chi_r-\sum_{k=1}^{i-1}\bar{\chi}_{k-1}
\chi_k}.\;\;\;\;\;\;\label{Gi} \ena

Of course, the expressions  for correlation functions  can be
caught  as well from the logarithmic derivatives of the partition
function $Z(B)$ [Eqs. (\ref{zh}) and (\ref{zi})] with respect to
inhomogeneous field $B(i,j)$, taken at
 $B(i,j)=0$.

{\centering
\subsection{Limit of an infinite lattice and large
distances.
 Magnetization
}}

In the limit of an infinite lattice, $N\to \infty$, one can
replace the sums in Eq. (\ref{j12}) by integrals in accordance
with Eq. (\ref{si}). After evaluation of the integral over $q$,
one will obtain
\begin{widetext}
 \bea \label{mcG}
&&\mathcal{G}_{2i\;2j\!+\!1}=\delta_{i,j}\!+\!
\frac{2}{\pi}\int_0^{\pi/2}
\frac{\cos{[2(i-j-1)p]}\sinh{[J_1]}\sinh{[J_2]}-\cos{[2(i-j)p]}}
{\sqrt{1\!+\!\left(\sinh{[2J_1]}\sinh{[2J_2]}\right)^2-2\cos{[2p]}\;\sinh{[2J_1]}\sinh{[2J_2]}}}\;
\mathrm{d}p,\nn\\ &&
\mathcal{G}_{2i\;2j}=\mathcal{G}_{2i\!+\!1\;2j\!+\!1}=0.\ena
%
The last expression in Eq. (\ref{mcG}) shows, that as in the
homogeneous case, on the infinite lattice in the inhomogeneous
case $J_1\neq J_2$ also we  can use determinant representation
(\ref{Gi1}) instead of Eq. (\ref{gint}),
\bea G(i)\! =\int \! D\bar{\chi}D\chi {\; e
\;}^{\sum_{k,r=0}^{i-1,i-1}
\bar{\mathcal{G}}'_{k-r\!+\!1}\bar{\chi}_k
\chi_r}=\mathrm{Det}[\mathcal{G}'(i)]\label{Gi2},\;\;\;\;\;\;\;\\\nn
[\mathcal{G}'(i)]_{k,\;r}\equiv
\bar{\mathcal{G}}'_{k\!+\!1-r},\qquad k,r=0,i-1,\ena
with
%
\bea\bar{\mathcal{G}}'_{k-r+1}=\frac{2}{\pi}\int_0^{\pi/2}
\frac{\cos{[2(k-r-1)p]}\sinh{[J_1]}\sinh{[J_2]} -\cos{[2(k-r)p]}}
{\sqrt{1\!+\!\left(\sinh{[2J_1]}\sinh{[2J_2]}\right)^2-2\cos{[2p]}\;\sinh{[2J_1]}\sinh{[2J_2]}}}\;
\mathrm{d}p. \label{gg}\ena
%
It is easy to see that the integral for the matrix elements in Eq.
(\ref{gg})  can be transformed into the form
\bea\nn && \bar{\mathcal{G}}'_{n\!+\!1}
\!=\!\frac{2}{\pi}\int_0^{\pi/2}\!\frac{\cos{[2(n\!-\!1)p]}\sinh{[2J_1]}\sinh{[2J_2]}\!-\!\cos{[2
n p]}}
{\sqrt{1\!+\!\left(\sinh{[2J_1]}\sinh{[2J_2]}\right)^2\!-\!2\cos{[2p]}\;\sinh{[2J_1]}\sinh{[2J_2]}}}\;
\mathrm{d}p\nn\\&&
\!=\!\frac{1}{\pi}\int_0^{\pi}\frac{\cos{[(n\!-\!1)p]}\sinh{[2J_1]}\sinh{[2J_2]}\!-\!\cos{[n
p]}}
{\sqrt{1\!+\!\left(\sinh{[2J_1]}\sinh{[2J_2]}\right)^2\!-\!2\cos{[p]}\;\sinh{[2J_1]}\sinh{[2J_2]}}}\;
\mathrm{d}p\nn\\&& \!=\!\frac{1}{2\pi}\int_0^{\pi}\!\frac{e^{i\;n
p}(e^{\!-\!ip}\sinh{[2J_1]}\sinh{[2J_2]}\!-\!1)\!+\!e^{\!-\!i\;n
p}(e^{ip}\sinh{[2J_1]}\sinh{[2J_2]}\!-\!1)}
{\sqrt{(e^{ip}\;\sinh{[2J_1]}\sinh{[2J_2]}\!-\!1)(e^{\!-\!ip}\;\sinh{[2J_1]}\sinh{[2J_2]}\!-\!1)}}\;
\mathrm{d}p\nn\\
&&\!=\!\frac{1}{2\pi}\int_{\!-\!\pi}^{\pi}\! e^{i(n
p)}\frac{\sqrt{{e^{\!-\!ip}\sinh{[2J_1]}\sinh{[2J_2]}\!-\!1}}}
{\sqrt{e^{ip}\;\sinh{[2J_1]}\sinh{[2J_2]}\!-\!1}}\; \mathrm{d}p\;
. \label{gc} \ena
\end{widetext}

It is well known, that one can investigate the magnetization
$\langle \sigma_1(i,j)\rangle$  by analyzing the large distance
asymptotes of two spin-correlation function on an infinite
lattice. Namely,
\bea (\langle
\bar{\sigma}_{\alpha}(i,j)\rangle)^2=\lim_{K\rightarrow \infty}
(\lim_{N \rightarrow \infty}\langle
\bar{\sigma}_\alpha(0,0)\bar{\sigma}_{\alpha'}(K,K)\rangle)\label{s}\\\nn
=\lim_{K\rightarrow \infty} (\lim_{N \rightarrow \infty}\langle
\bar{\sigma}_{\alpha}(0,0)\bar{\sigma}_{\alpha'}(0,K)\rangle),
\ena
where $N$ is the linear size of the square lattice.

In the Ref. \cite{MV} it was shown that spin-spin correlation
functions
$\langle\bar{\sigma}_{\alpha}(0,0)\bar{\sigma}_{\alpha'}(i,i)\rangle$
and
$\langle\bar{\sigma}_{\alpha}(0,0)\bar{\sigma}_{\alpha'}(0,i)\rangle$
(for $T<T_c$) have a determinant representation. These correlation
functions have been represented as a determinant of an $i\times i$
matrix, $\mathcal{C}_i$, of the Toeplitz type,
\bea \label{Toep}\mathcal{C}_i=\left(\ba{cccc}c_0&c_{-1}&\cdots&c_{-i\!+\!1}\\
c_1&c_{0}&\cdots&c_{-i+2}\\
\vdots &\vdots &\vdots
&\vdots\\c_{i-1}&c_{i-2}&\cdots&c_{0}\ea\right).\ena In Eq.
(\ref{Toep}) the matrix elements are given by
 \bea
c_{n}=\frac{1}{2 \pi}\int_0^{2\pi}d\theta e^{-i n
\theta}\mathrm{C}(e^{i \theta}),\\\nn \mathrm{C}(e^{i
\theta})=\left(\frac{(1-\alpha_1 e^{i\theta})(1-\alpha_2
e^{-i\theta})}{(1-\alpha_1 e^{-i\theta})(1-\alpha_2
e^{i\theta})}\right)^{1/2}. \ena
For the case
\bea\lim_{N\to
\infty}\langle\bar{\sigma}_{\alpha}(0,0)\bar{\sigma}_{\alpha'}(i,i)\rangle=
\mathrm{Det}[\mathcal{C}_i],\ena
$(\alpha_i)$'s  are defined as follows:
 \bea
\alpha_1=0,&\quad &
\alpha_2=(\sinh{[2J_1]}\sinh{[2J_2]})^{-1}.\ena

Careful analysis of the matrix $\mathcal{G}'(i)$, given by Eq.
(\ref{Gi2}), shows, that due to Eqs. (\ref{gg} and \ref{gc}) and
after some rearrangement of its rows, which leave the determinant
invariant, $\mathcal{G}'(i)$ coincides with $\mathcal{C}_i$.
 Note that in our notations the coordinate plane on the lattice
 is $45^0$ rotated  with respect to the coordinate plane in Ref. \cite{MV}, so the
 correlation function
 of the spins arranged
in the horizontal or vertical lines in our case coincide with
${\langle\bar{\sigma}_{\alpha}(0,0)\bar{\sigma}_{\alpha'}(i,i)\rangle}$
derived in Ref. \cite{MV}.

Now one can follow the technique developed in Ref.\cite{MV}, based
on the Szeg\"{o}'s theorem, and find the solution for the
magnetization. The theorem can be applied when $T<T_c$ and
directly reproduces the known result for the magnetization
\cite{O,Y,Baxter} originally derived by Yang in article \cite{Y}:
\bea\lim_{i\rightarrow\infty}{\langle
\bar{\sigma}_{\alpha}(0,0)\bar{\sigma}_{\alpha'}(0,i)\rangle}\!\!\!&=&\!\!\!
\lim_{i\rightarrow\infty}{\langle
\bar{\sigma}_{\alpha}(0,0)\bar{\sigma}_{\alpha'}(i,i)
\rangle}\label{snh}\\\nn=
\left(\frac{(1\!-\!\alpha_2^2)(1\!-\!\alpha_1^2)}{(1\!-\!\alpha_1\alpha_2)^2}\right)^{\frac{1}{4}}
\!\!\!&=&\!\!\!\left(1\!-\!(\sinh{[2J_1]}\sinh{[2J_2]})^{\!-\!2}\right)^{\frac{1}{4}}.\ena

{\centering
\section{Related one-dimensional quantum problem}}


 It is possible to connect the partition function of the
quantum 1DIM to the partition function of two-dimensional
classical system (2DIM) using limit (\ref{trot}) and the Trotter
formula, see Ref. \cite{C}.

As it was stated in the second section, the transfer matrix of
two-dimensional model, which is defined as a product of $R$
matrices, plays a role of the discrete time evolution operator
defined on a 1D chain.

In this section we investigate the transfer matrix given by Eq.
(\ref{tr1}) and express it via one-dimensional fermionic fields
defined on a chain. By the convention, the trace of the transfer
matrix can be connected with the partition function of the quantum
chain model, defined by Hamiltonian operator $\mathcal{H}$,
$$\mathrm{tr}\;\tau=\mathrm{tr}\;e^{-\mathcal{H}}.$$
 The trace in
the definition of the transfer matrix,
$\tau_j=\mathrm{tr}_1{\prod_{i}{R}(i,j)} $, in Eq. (\ref{tr1}) is
taken over the variables which have even-even lattice coordinates
(denoted by black circles on the figures); ${R}(i,j)$ matrices are
arranged along the horizontal chain with $N$ vertices (white
circles on the figures). In the following we shall omit the
coordinate indices and will use only indices denoting the vertices
on the chain. Using $R$ operators represented in terms of
fermionic creation-annihilation operators,
$\mathcal{R}(\mathbf{c}^+_1,\mathbf{c}_1;c^+_2,c_2)$ [Eqs. (\ref{R
fer}) and (\ref{rfg})], one easily comes to the transfer matrix
\bea \tau(\{c^+_n,c_n\})=
\mathrm{tr}_{1}{\prod_{i=1}^N{\mathcal{R}}(\mathbf{c}^+_1,\mathbf{c}_1;c^+_i,c_i)}.
\label{tc} \ena
We can evaluate the trace in Eq. (\ref{tc}) passing to the
coherent basis with Grassman variables
for the fermionic operators $\mathbf{c}^+_1,\mathbf{c}_1$ and
$\{c^+_i,c_i\}$. After integration by the variables corresponding
to the operators $\mathbf{c}^+_1,\mathbf{c}_1$, we shall arrive at
[we have chosen the homogeneous case $b=b',\;c=c',\;d=d'$,
(\ref{rfg})]
\bea t(\{\bar{\psi},\psi\})=
\prod_i\langle\bar{\psi}_i|\tau(\{c^+_n,c_n\})\prod_i|\psi_i\rangle\\\nn
= ({R_{00}^{00}}^{N}+{R_{01}^{10}}^N)
e^{\!-\!\mathbb{H}(\bar{\psi}_i,\psi_i)}\label{tc1},\ena
 \bea
&&\!-\!\mathbb{H}(\bar{\psi}_i,\psi_i)\label{acd}\\\nn && =
\sum_{k=1}^{N}\frac{(\!-\!1)^kc^{N\!-\!k}\Delta^k}{(1+c^N)^k}\sum_{i_1<\cdots
<i_k} {\textsl{n}_{i_1}\cdots\textsl{n}_{i_k}} +c
\sum_{i=1}^{N}\textsl{n}_i
\\\nn&&+\frac{1}{1+c^N}\left(
1+\sum_{k=1}^{N}\frac{(\!-\!1)^k}{(1+c^N)^k}\left[\prod_{i=1}^N
(c\!-\!\Delta\textsl{n}_i)\!-\!c^N\right]^k\right)\\ &&\times
\sum_{i,j}\left(b\;\bar{\psi}_i\!-\!d\;{\psi}_i\right)
K_{i,j}\left(b\;\psi_{j\!-\!1}+d\;\bar{\psi}_{j\!-\!1}\right)\nn\\
&&K_{i,j}\\
&&\nn\!=\!\left\{\!\ba{cc}1&i=j\\
(c\!-\!\Delta \textsl{n}_j)\cdots(c\!-\!\Delta \textsl{n}_{i\!-\!1})&i>j\\
\!-\!(c\!-\!\Delta \textsl{n}_1)\!\cdots\!(c\!-\!\Delta
\textsl{n}_{i\!-\!1}) (c\!-\!\Delta
\textsl{n}_j)\!\cdots\!(c\!-\!\Delta \textsl{n}_{N})&i<j\ea\right.
 \ena
Here $\textsl{n}_i=\bar{\psi}_i\psi_i$.
Correspondingly the normal ordered expression of $\tau(\{c^+_n,c_n\})$ is
\be\tau(\{c^+_n,c_n\})=({R_{00}^{00}}^{N}+{R_{01}^{10}}^N):
e^{-\mathbb{H}(c^+_i,c_i)-\sum_i c^+_i c_i}:.
\ee
 The expression of
$\mathbb{H}$ in (\ref{acd}) simplifies if $c=0$. In case of
$\Delta=0$ the function $\mathbb{H}$ is a quadratic function and
admits diagonalization by means of Fourier transformation.
 \paragraph*{IM, XY.} Here we are presenting the transfer matrices, which
 correspond to the free-fermionic cases: $\Delta=0$ in Eq. ({\ref{rfg}}),
 i.e.,
 IM [Eq. (\ref{r})]
 and $XY$ model [Eqs. (\ref{rxyz} and \ref{ii'})]. Now logarithm of Eq. (\ref{tc})
is a quadratic function  over  $N$ pairs of fermion operators,
$\{c^+_n,c_n\}$, due to the Eqs. (\ref{rf} and  \ref{rf1}).
After performing
Fourier transformation
 for operators $c^+_n, c_n$ in Eq. (\ref{acd}), the transfer matrix takes the form
\bea \label{ccd}&& \tau =(R_{00}^{00})^{N}
\left(1\!-\!c^N\right) :\exp\{\sum_{p=0}^{N/2\!-\!1}
\mathbb{H}(p)\}:; \\
&&\mathbb{H}(p)=\left(c\!-\!1\!+\!\frac{b^2e^{i\pi\frac{2p\!+\!1}{N}}}{1\!-\!c\;e^{i
\pi\frac{2p\!+\!1}{N}}} \!+\!
\frac{d^2e^{\!-\!i\pi\frac{2p\!+\!1}{N}}}{1\!-\!c\; e^{\!-\!i
\pi\frac{2p\!+\!1}{N}}}\right)c^+_p c_p
\nn\\
&&\!+\!\left(c\!-\!1\!+\!\frac{b^2e^{\!-\!i\pi\frac{2p\!+\!1}{N}}}{1\!-\!c\;e^{\!-\!i
\pi\frac{2p\!+\!1}{N}}} \!+\!
\frac{d^2e^{i\pi\frac{2p\!+\!1}{N}}}{1\!-\!c\;e^{i
\pi\frac{2p\!+\!1}{N}}}\right)c^+_{N\!-\!p\!-\!1}c_{N\!-\!p\!-\!1\nn}\\
&&\!+\!\frac{2\;i\;b \;d \; \sin\left[
\pi\frac{2p\!+\!1}{N}\right]}{1\!+\!c^2\!-\!2\;
c\;\cos\left[\pi\frac{2p\!+\!1}{N}\right]}(c^+_p
c^+_{N\!-\!p\!-\!1}\!+\! c_p\; c_{N\!-\!p\!-\!1}).\ena
%

In the course of calculation of the partition function in Sec. III
we have diagonalized this type of quadratic expression by a simple
change of basis  (\ref{rep}). Recall that here $c^+_p, c_p$ are
not Grassmann variables but rather fermionic operators and {\em
any transformation must keep anticommutation relations}. So we
distinguish two kind of fermion fields, defined as $c_{\alpha
\;p}$, $\alpha=1,2$,
 \bea
c^+_{1p}=c^+_p,\; c_{1p}=c_p,\;c^+_{2p}=c_{N-p-1},\;
c_{2p}=c^+_{N-p-1}.\ena
These replacements bring the operator $\sum_{p=0}^{N/2-1} 
\mathbb{H}(p)$ to
the form
\be\sum_{p=0}^{N/2-1} \sum_{\alpha,\;\beta=1,\;2} 
\mathbb{H}'_{\alpha\;
\beta}(p)c^+_{\alpha\;p} c_{\beta\;p}.\ee
 The task now is to
diagonalize the matrix
\be
\label{ccb}
\mathbb{H}'(p)=\left( \ba{cc}r_1(p)&r_2(p)\\
-r_2(p)&-r_1(-p)\ea \right),\ee
where
 \bea
r_1(p)=c-1+\frac{b^2e^{i\pi\frac{2p+1}{N}}}{1-c\; e^{i
\pi\frac{2p+1}{N}}} + \frac{d^2\;e^{-i\pi\frac{2p+1}{N}}}{1-c\;
e^{-i \pi\frac{2p+1}{N}}},\;\;\quad\nn
\\
r_2(p)=\frac{ 2\; i\;b\; d\; \sin\left[
\pi\frac{2p+1}{N}\right]}{1+c^2-2
c\;\cos\left[\pi\frac{2p+1}{N}\right]}.\qquad \ena
We can represent the transfer matrix given by Eq. (\ref{ccd}) in
the following diagonal form:
\bea \tau\approx e^{\sum_{p=0}^{N/2-1} \left( a'_{+}(p)c'^+_{1\;p}
c'_{1\;p}+ a'_{-}(p)c'^+_{2\;p} c'_{2\;p}\right)}. \ena
with the eigenvalues of matrix (\ref{ccb}),
\be a'_{\pm}(p)\!=\! \frac{1}{2}\left(\!r_1(p)\!-\!r_1(-p)\!\pm\!
\sqrt{(r_1(p)\!+\!r_1(-p))^2\!-\!4(r_2(p))^2}\right).
\label{eig}\ee
Thus, we arrive at a 1D quantum system defined with Hamiltonian
operator
\bea \mathcal{H}=-\sum_{p=0}^{N/2-1} \left( a'_{+}(p)c'^+_{1\;p}
c'_{1\;p}+ a'_{-}(p)c'^+_{2\;p} c'_{2\;p}\right). \label{hxz}\ena
Particularly, for the IM [where $b,c,d$ are defined as in Eq.
(\ref{abc})], in the homogeneous case, $J_1=J_2=J$, we have
$r_1(p)=r_1(-p)$ and eigenvalues (\ref{eig}) acquire the form
\be a'_{\pm}(p)= \pm \sqrt{|r_1(p)|^2+|r_2(p)|^2}. \ee
%
%
%
The ground state of the system is composed by the negative-energy
modes. In the thermodynamic limit, $N\rightarrow \infty$, the gap
between two spectral curves, $a'_{\pm}(p)$, is found at the Fermi
points with momenta $0,\; \pi$ and is equal to
\be [a'_{+}(0)\!-\!a'_{-}(0)]\equiv 2 r_1(0)=
2\left(\!1\!-\!\sinh[\frac{2J}{T}]\!\right)\left(\!1\!+\!\sinh[\frac{2J}{T}]\!\right)\!.
\ee
We see that $r_1(0)$ vanishes at the critical temperature $T_c$ of
2DIM, given by
  $\sinh[2J/T_c]=1$, as
 \be
[a'_{+}(0)-a'_{-}(0)]\sim(T-T_c),\label{at}\ee
 demonstrating that
at $T=T_c$
 the 1D system is gapless and has no massive excitations.
 Behavior (\ref{at}) holds true for the inhomogeneous case $J_1\neq J_2$
 also.
\section{Summary}

In this work we have presented an approach to the investigation of
two-dimensional statistical models, basing on the fermionic
formulation of the vertex $R$ matrices (Boltzmann weights). If the
operator form of the $R$ matrix in terms of scalar fermionic
creation and annihilation operators has definite even grading [for
$XYZ$ model and 2DIM see Eq. (\ref{R fer})], then fermionic
representation of $R(i,j)$ on the lattice acquires local
character. If the operators have indefinite grading
 [models in the presence of an external magnetic
field, see Eq. (\ref{rH})], then one must take into account
Jordan-Wigner non-local operator, as in Eq. (\ref{zh}), which is
discussed in details in the Appendix.

For the models under consideration we derive partition functions
as continual integrals with corresponding field theoretical
actions on the square lattice: Eq.  (\ref{AA}) gives the fermionic
action corresponding to the general eight-vertex model, which
includes both $XYZ$ model and two-dimensional Ising model.
Although there is a correspondence between 2DIM and $XZ$ models,
we straightforwardly presented the  $R$ matrix of the 2DIM in Eq.
(\ref{rt}) as a solution of  Yang-Baxter equation which ensures
the integrability of the model. For the free-fermionic case  the
direct calculation of the partition function and  correlation
functions is
performed [Eqs (\ref{det}) and (\ref{skz}]. In case of the 2DIM
the continuum limit of the two-dimensional action is presented in
Eq. (\ref{contin}) and the known thermodynamic and magnetic
characteristics are reproduced [see Eqs. (\ref{DET}), (\ref{OO})
and (\ref{snh})]. We also consider  2DIM in the presence of a
finite magnetic field and corresponding nonlocal fermionic action
is evaluated [Eq. (\ref{zi})].

In  light of  correspondence of two-dimensional classical
statistical models and one-dimensional quantum models  we obtain
one-dimensional quantum fermionic Hamiltonian operator (\ref{acd})
for
eight-vertex model. 
For  free-fermionic cases the Hamiltonian operators are brought
to the diagonal form (\ref{hxz}), the spectral analysis  of which
reflects the critical behavior of the underling models.

\section*{Acknowledgements}

Sh.~Kh. thanks the Volkswagen Foundation for the partial financial
support.


{\centering
\section*{Appendix}}
%
\setcounter{section}{0}\setcounter{footnote}{0}
\addtocounter{section}{0}\setcounter{equation}{0}
\renewcommand{\theequation}{$\textmd{A}$.\arabic{equation}}

\paragraph*{Jordan-Wigner transformation.
}

Fermionic representation of spin states naturally introduces
grading for both states and operators. $\bar{\sigma}_{\alpha},\;
\alpha=0,1$  spin states can be represented by
$|0\rangle,\;|1\rangle$ fermionic states with zero and one
fermions. Single fermion states are anticommuting at different
points of the lattice.  The same property takes place for the odd
operators in terms of fermionic creation and annihilation
operators. This property does not hold for spin states and
operators. Therefore, if one would like to represent the action of
odd number of spin operators $\sigma_1^{(k)}$ defined in the space
of spins (nongraded space)
%
\bea \{\hat{1}^{(1)} \otimes \hat{1}^{(2)}\! \cdots\! \otimes
\sigma_1^{(k)}\otimes\cdots\otimes
\hat{1}^{(n)}\}\!:\!|\alpha_1\rangle|\alpha_2\rangle\!\cdots\!
|\alpha_n\rangle, {\;\;\;}\ena
%
in terms of fermionic operators $(c+c^+)^{(k)}$, which act on
graded states $|\alpha_k\rangle$, one has to take into account the
graded behavior of all states $|\alpha_i\rangle$, $i<k$, placed
before the state $|\alpha_k\rangle$. This can be done
with the help of the operator $1-2n$, action of which on the state
$|\alpha\rangle$ depends on the parity, $p(\alpha)=\alpha$, as
follows:
\be (1-2n)|\alpha\rangle=(-1)^{p(\alpha)}|\alpha\rangle.
\ee
Using these operators, one can represent the action of a spin
operator $\sigma_1^{(k)}$, as
%
\bea\{\hat{1}^{(1)} \otimes \hat{1}^{(2)} \cdots \otimes
\sigma_1^{(k)}\otimes\cdots\otimes \hat{1}^{(n)}\}\nn
\\\Rightarrow (c+c^+)^{(k)}(1-2 c^+ c)^{(1)} \cdots(1-2 c^+ c
)^{(k-1)}.\label{sigma} \ena
This expression constitutes the inverse Jordan-Wigner spin-fermion
nonlocal transformation.

It is clear that for the product of two odd operators at different
points one needs to take into account only the states between
them,
\bea\nn &\{\cdots\hat{1}^{(i-1)}\otimes\sigma_1^{(i)} \otimes
\hat{1}^{(i+1)}\!\cdots\! \hat{1}^{(k-1)}\otimes\sigma_1^{(k)}
\otimes \hat{1}^{(k+1)}\!\cdots \}
& \\ \label{JW} &\Rightarrow (c+c^+)^{(i)}\prod_{r=i}^{k-1}(1-2
c^+ c)^{(r)}(c+c^+)^{(k)},& \ena
which is a consequence of the property
\be(1-2 c^+ c )^{(i)}(1-2 c^+ c )^{(i)}=1.\label{n1} \ee
Note, that operator $(1-2 c^+ c )$ is the fermionic form
corresponding to the Pauli matrix $\sigma_z$. This means that if
we place the operators $\sigma_z^{(i)}$ instead of unity $1^{(i)}$
in Eq. (\ref{sigma}) for all $i<k$, we shall have
\bea \{\sigma_z^{(1)} \otimes \sigma_z^{(2)} \cdots \otimes
\sigma_1^{(k)}\otimes\cdots\otimes \hat{1}^{(n)}\}\Rightarrow
(c+c^+)^{(k)}.{\;\;\;}\ena
Similarly, we have
\bea \label{1z1} \{\ldots
\hat{1}^{(i-1)}\otimes(\sigma_1\sigma_z)^{(i)} \otimes
\sigma_z^{(i+1)}\ldots\nn\\
\ldots \sigma_z^{(k-1)}\otimes\sigma_1^{(k)} \otimes
\hat{1}^{(k+1)} \ldots  \}\\\nn\Rightarrow
(c+c^+)^{(i)}(c+c^+)^{(k)}.\quad \ena

 \paragraph*{Jordan-Wigner spin-fermion transformation on
the two-dimensional lattice.}

In Sec. II  the partition function Eq. (\ref{tr2}) was defined as
an expectation value of the products of $R$ operators. These
products can be rewritten as
\bea\label{zsigma}
Z=\sum_{\{\alpha_{2i+1,1}\}_{i=0,N-1}}\sum_{\{\alpha_{0,2j}\}_{j=1,N}}\\\nn
\langle\Sigma|\;\prod_{j=N}^{1}\prod_{i=N-1}^{\;0}R(i,j)
\;\;|\Sigma\rangle,\ena
where the trace is taken over both "auxiliary" and "quantum"
states,
$$|\Sigma\rangle=|\alpha_{0,2N}\rangle\cdots|\alpha_{0,4}\rangle|\alpha_{0,2}\rangle
|\alpha_{1,1}\rangle|\alpha_{3,1}\rangle\cdots
|\alpha_{2N-1,1}\rangle.$$

In fermionic representation described in Sec. III, the states
$|\alpha_{i,j}\rangle$ acquire grading and the arrangement in
$|\Sigma\rangle$ becomes significant. Fermionic $R$ operator given
by Eq. (\ref{R fer}) has zero parity, which ensures the local
"fermionization" of the partition function: each $R$ operator in
Eq. (\ref{zsigma}) can be replaced with its fermionic counterpart
without any "tail". But the formulas of spin-spin correlation
functions contain the spin operator $\sigma_1(k,r)$, which in the
fermionic formulation has odd parity. From the inverse
Jordan-Wigner transformation in Eq. (\ref{JW}) it follows that
 the fermionic operator corresponding
to $\sigma_1(k,r)$ should contain non-local operator
$\prod(1-2n(i,j))$,
 where the product runs over sites $(i,j)$, arranged before the
 site
 $(k,r)$.

 Recall, that the mean value of the
operator $\sigma_1(2k,2r)$ is defined by
\bea \label{sr}
\langle\sigma_1(2k,2r)\rangle=\frac{1}{Z}\sum\qquad\qquad\qquad\qquad\\
\langle\Sigma|\left(R\cdots
 R(k,r) \sigma_1(2k,2r) R(k-1,r) \cdots
 R\right)|\Sigma\rangle. \quad\nn
 \ena
 And due to the conventions adopted in the previous sections, the
$R(i,j)$ operator acts as
\begin{eqnarray}
\label{RIJ} R(i,j)|\alpha_{2i,2j}\rangle|\alpha_{2i+1,2j-1}\rangle
\quad\quad \quad{\;\;\;} \\\nn =
R_{\;\;\;\;\alpha_{2i,2j}\;\;\;\;\;\alpha_{2i+1,2j-1}}^{\alpha_{2i+1,2j+1}\;\alpha_{2i+2,2j+2}}
|\alpha_{2i+1,2j+1}\rangle|\alpha_{2i+2,2j+2}\rangle,\quad
\end{eqnarray}
with the matrix elements defined by Eq.  (\ref{elements}). Then
one can notice, that the action of $R$ operators, placed on the
right side of $\sigma_1(2k,2r)$ in the right hand side of Eq.
(\ref{sr}), on the state $|\Sigma\rangle$, transforms it to the
following state:
\bea \prod_{i=k-1}^{0} R(i,r)\prod_{j=r}^{1}\prod_{i=N-1}^{0}
R(i,j)|\Sigma\rangle\quad {\;\;\;} {\;\;\;}
\\\nn
\Rightarrow
|\alpha_{0,2N}\rangle\cdots|\alpha_{0,2r\!+\!2}\rangle|\alpha_{1,2r\!+\!1}\rangle\cdots
{\;\;\;}{\;\;\;}\\\nn
|\alpha_{2k-1,2r\!+\!1}\rangle|\alpha_{2k,2r}\rangle|\alpha_{2k\!+\!1,2r-1}\rangle\cdots
|\alpha_{2N-1,2r-1}\rangle.\ena
Hence, according to Eq. (\ref{sigma}), operator $\sigma_1(2i,2j)$
in its fermionic formulation reads
\bea
\label{s2k2r}
[c(2k,2r)\!+\!c^+(2k,2r)]\prod_{i=k-1}^{0}\!(1\!-\!2n(2i\!+\!1,2r\!+\!1))\nn\\
\times \prod_{j=r\!+\!1}^{N}\!(1-2n(0,2j)).\qquad \qquad
\ena
Similarly, in expression for the vacuum average value of spin
operators $\sigma_1(2k\!+\!1,2r\!+\!1)$, defined at odd-odd sites,
\bea
\label{sr1}
 \langle\sigma_1(2k\!+\!1,2r\!+\!1)\rangle=
\frac{1}{Z}\sum\qquad \qquad \qquad \qquad \\\nn
\langle\Sigma|\left(R\!\cdots\!
 R(k\!+\!1,r)\sigma_1(2k\!+\!1,2r\!+\!1) R(k,r)\! \cdots\! R\right)|\Sigma\rangle,
 \ena
the $R$ operators on the right-hand side of $\sigma_1(2k+1,2r+1)$
transform the state $|\Sigma\rangle$ into
 \bea \prod_{i=k}^{0}
R(i,r)\prod_{j=r}^{1}\prod_{i=N-1}^{0}
R(i,j)|\Sigma\rangle\Rightarrow\qquad\\\nn
|\alpha_{0,2N}\rangle\cdots|\alpha_{0,2r\!+\!2}\rangle|\alpha_{1,2r\!+\!1}\rangle\cdots\qquad\\\nn
|\alpha_{2k\!+\!1,2r\!+\!1}\rangle|\alpha_{2k\!+\!2,2r}\rangle|\alpha_{2k\!+\!3,2r-1}\rangle\cdots
|\alpha_{2N-1,2r-1}\rangle,\ena
which means that in its fermionic formulation
$\sigma_1(2k+1,2r+1)$ is equipped with the same non-local operator
as Eq. (\ref{s2k2r})
\bea\label{s2k12r1}
[c(2k+1,2r+1)+c^+(2k+,2r+1)]\qquad\\
\nn \prod_{i=k-1}^{0}(1\!-\!2n(2i\!+\!1,2r\!+\!1))
\prod_{j\!=\!r\!+\!1}^{N}(1\!-\!2n(0,2j)).\ena
As an example, consider the spin operators on the vertices $(5,3)$
and $(6,2)$ in Fig. \ref{fig2}. There the positions of the $1-2n$
operators are marked by arrows at the corresponding sites. If spin
operators are placed on the edges of the lattice,
  $\sigma_1(2i+1,1)$ and
$\sigma_1(0,2j)$, they immediately act on $|\Sigma\rangle$ and can
be replaced by the fermionic operators,
$$[c(2i\!+\!1,1)\!+\!c^{\!+\!}(2i\!+\!1,1)]\prod_{r=1}^{N}(1\!-\!2n(0,2r))\prod_{k=1}^{i-1}(1\!-\!2n(2k\!+\!1,1))
$$
 and
$$[c(0,2j)+c^{+}(0,2j)]\prod_{r=j+1}^{N}(1-2n(0,2r)),$$
 respectively, in accordance with general expressions in Eqs.
 (\ref{s2k2r}) and
 (\ref{s2k12r1}).

In order to calculate the correlation function $\langle
\sigma_1(i,j)\sigma_1(k,r)\rangle$
 in the fermionic operator form, it is necessary  to
replace $\sigma_1(i,j)$ by corresponding fermionic operators
(\ref{s2k2r}) and (\ref{s2k12r1}). As it is shown in the first
part of this section, coinciding operators $1-2n$  in fermionic
counterparts
 of $\sigma_1(i,j),\;\sigma_1(k,r)$ operators
cancel each other and only operators placed on a path, which
connects points $(i,j)$ and $(k,r)$, will be left.
 The choice of the path is
arbitrary, which is a result of property (\ref{n1}),
$1-2n=\sigma_z$  and
\be
(\sigma_z\otimes\sigma_z)\;\mathbf{R}\;(\sigma_z\otimes\sigma_z)
=\mathbf{R},\label{n2} \ee
%
%
%
with $\mathbf{R}$ operator defined in Eq. (\ref{R1}).

 More precisely, a fermionic realization for the
product of two spin operators, $\sigma_1(i,j)$ and
$\sigma_1(k,r)$, when $i<k,\;j<k$, has the form presented in Eq.
(\ref{cc}). It looks as if one inserts into the vertices on a path
between $(i,j)$ and $(k,r)$ points, operators $\sigma_z\;(=1-2n)$,
instead of unity operators in the spin representation and vice
versa: it is a simple task to derive the correlation function
$\langle
\sigma_1(i,j)\left(\prod\sigma_z(i',j')\right)\sigma_1(k,r)\rangle$
on the two-dimensional lattice, where the operators $\sigma_z$ are
placed
 on a path of
vertices connecting points $(i,j)$ and $(k,r)$. It can be done by
replacing operators $\sigma_1$ by $(c+c^+)$ and finding
corresponding Green's functions
[see Eq. (\ref{1z1})].


\end{document}